\documentclass[aps,prb,twocolumn,superscriptaddress,longbibliography]{revtex4-2}
\usepackage{graphicx}
\usepackage{latexsym}
\usepackage{amssymb}
\usepackage{amsmath}
\usepackage{amsfonts}
\usepackage{upgreek}
\usepackage{float}
\usepackage{bm}
\usepackage{multirow}
\usepackage{color}
\usepackage[T1]{fontenc}
\usepackage{hyperref}
\usepackage{subfigure}
\usepackage{booktabs,tabularx}

\hypersetup{
colorlinks = true,
linkcolor = [rgb]{0.70,0.13,0.13},
citecolor = [rgb]{0.13,0.55,0.13},
urlcolor  = [rgb]{0.25, 0.41, 0.88}}

\newcommand{\rrangle}[1]{\left|{#1}\right\rangle}
\newcommand{\llangle}[1]{\left\langle{#1}\right|}

\begin{document}
\title{Many-body phase transitions in a non-Hermitian Ising chain}
\author{Chao-Ze Lu}
\affiliation{College of Physics, Nanjing University of Aeronautics and Astronautics, Nanjing, 211106, China}
\affiliation{Key Laboratory of Aerospace Information Materials and Physics (NUAA), MIIT, Nanjing 211106, China}

\author{Xiaolong Deng}
\affiliation{Leibniz-Rechenzentrum, Boltzmannstr. 1, D-85748 Garching bei M\"{u}nchen, Germany}

\author{Su-Peng Kou}
\affiliation{School of Physics and Astronomy, Beijing Normal University, Beijing 100875, China}

\author{Gaoyong Sun}
\thanks{Corresponding author: gysun@nuaa.edu.cn}
\affiliation{College of Physics, Nanjing University of Aeronautics and Astronautics, Nanjing, 211106, China}
\affiliation{Key Laboratory of Aerospace Information Materials and Physics (NUAA), MIIT, Nanjing 211106, China}

\begin{abstract}
We study many-body phase transitions in a one-dimensional ferromagnetic transversed field Ising model with an imaginary field and show that
the system exhibits three phase transitions: one second-order phase transition and two $\mathcal{PT}$ phase transitions. The second-order phase transition occurring in the ground state is investigated via biorthogonal and self-normal entanglement entropy, for which we develop an approach to perform finite-size scaling theory to extract the central charge for small systems. Compared with the second-order phase transition, the first $\mathcal{PT}$ transition is characterized by the appearance of an exceptional point in the full energy spectrum, while the second $\mathcal{PT}$  transition only occurs in specific excited states. Furthermore, we interestingly show that both of exceptional points are second-order in terms of scalings of imaginary parts of the energy. This work provides an exact solution for many-body phase transitions in non-Hermitian systems.


\end{abstract}

\maketitle

\section{Introduction}
Quantum phase transitions, which represent changes in the ground state of a system controlled by external parameters, are a fundamental concept in the field of condensed matter physics\cite{Sachdev1999}. Most quantum phase transitions can be understood via Ginzburg-Landau
Symmetry breaking\cite{ginzburg2009theory,tsuei2000pairing} or Wilson renormalization group theory\cite{Wilson1974,Wilson1975}. Phase transitions caused by symmetry breaking are usually characterized by order parameters\cite{Sachdev1999,tsuei2000pairing}. Finite-size scaling theory \cite{Fisher1972,Fisher1974} can be used as a powerful method to analyze phase transitions based on finite-system order parameters.

Thermal phase transitions (TPTs), caused by thermal fluctuations rather than quantum fluctuations, are well-known finite-temperature phase transitions \cite{Sachdev1999} in statistical mechanics. In addition to TPTs, dynamic quantum phase transition (DQPTs) \cite{heyl2013dynamical,heyl2018dynamical} and excited-state quantum phase transitions (ESQPTs) \cite{cejnar2006monodromy,cejnar2007coulomb,caprio2008excited,puebla2016excited,feldmann2021interferometric,cejnar2021excited} are two other quantum phase transitions beyond the ground state quantum phase transition. DQPT concerns the excited states of a system in its time evolution\cite{heyl2013dynamical,heyl2018dynamical}, but in contrast, ESQPT depends directly on the structure of the full spectrum\cite{cejnar2006monodromy,cejnar2007coulomb,caprio2008excited,puebla2016excited,feldmann2021interferometric,cejnar2021excited}.

\begin{figure}[t]
	\centering
	\includegraphics[width=8.7cm]{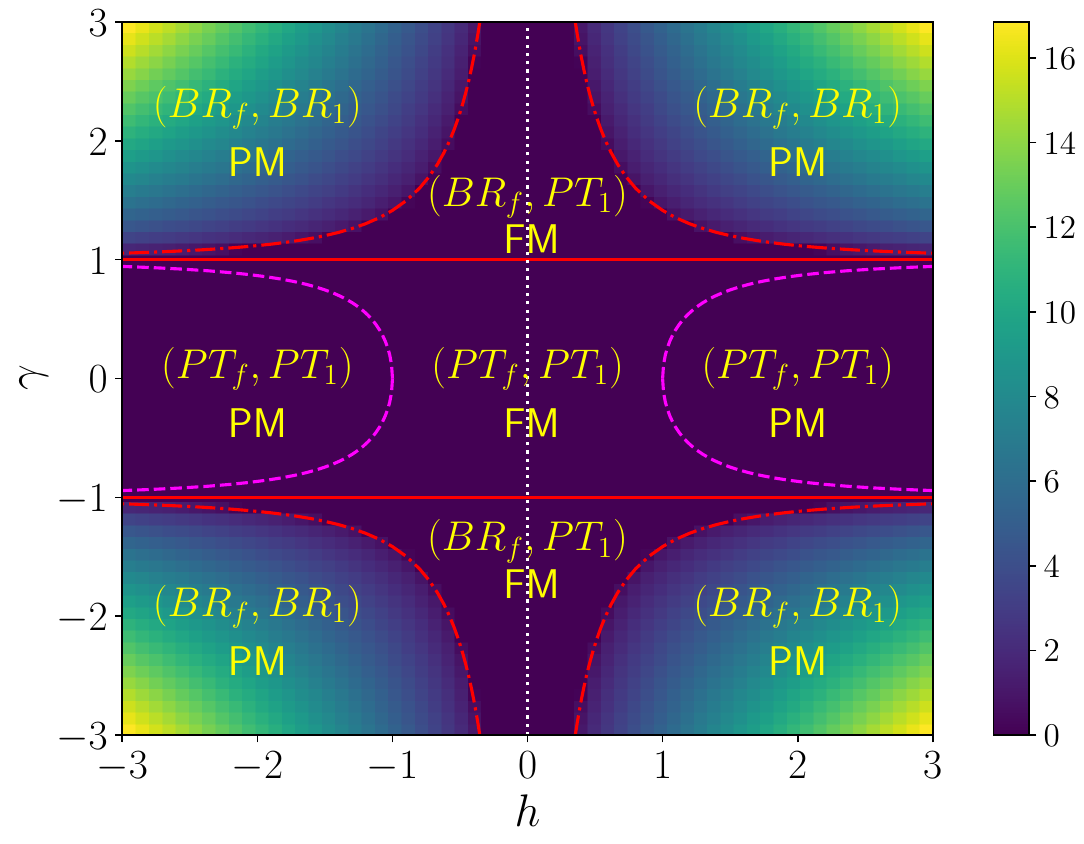}
	\caption{The phase diagram of the NHTI model with respect to $h$ and $\gamma$. 
	The ground-state, full $\mathcal{PT}$, first excited-state $\mathcal{PT}$ transitions are marked by the dashed magenta, red solid and red dash-dot lines, respectively.
	Here, the white dotted line indicates the system is Hermitian at $h=0$. 
	The color bar represents imaginary parts of first excited-state energies obtained with $L=8$ and $J=1$.
	The notations FM and PM denote ground-state phases,
	$PT_f$ and $BR_f$ represent $\mathcal{PT}$ symmetric and $\mathcal{PT}$ broken phases in full many-body energy spectra. 
	While $PT_1$ and $BR_1$ indicate $\mathcal{PT}$ symmetric and $\mathcal{PT}$ broken phases of first excited states.}
	\label{fig:PhaseDiagram}
\end{figure}

In non-Hermitian systems, the $\mathcal{PT}$ phase transition, which reveals the full energy spectrum as well \cite{bender1998real,bender2002complex}, is another fascinating phase transition that has attracted great interest in recent years \cite{bergholtz2021exceptional,ashida2021non}. As critical points of $\mathcal{PT}$ phase transition and a unique feature of non-Hermitian systems, many unknown properties of non-Hermitian exceptional points remain to be explored. Recently, it has been argued interestingly that phase transitions may occur in specific excited states, such as the first excited state rather than the ground state \cite{zhang2022symmetry,lu2023many,yang2023first}. For non-Hermitian systems, this phase transition corresponds to a $\mathcal{PT}$ transition between the first and second excited states \cite{zhang2022symmetry,lu2023many}. These $\mathcal{PT}$ phase transitions, dubbed as {\it first excited-state $\mathcal{PT}$ transitions}, are caused by nearest-neighbor interactions and strongly supported by numerical simulations \cite{zhang2022symmetry,lu2023many}. The understanding of first excited-state $\mathcal{PT}$ transitions is an interesting issue.

In the paper, we show that first excited-state $\mathcal{PT}$ transitions can also be caused by imaginary fields in a $\mathcal{PT}$ symmetric system. We provide exact solutions for the emergence of first excited state $\mathcal{PT}$ transitions in a one-dimensional non-Hermitian ferromagnetic transverse field Ising (NHTI) model. Furthermore, we show that both the first excited-state and the full $\mathcal{PT}$ transitions belong to second-order exceptional points from the finite-scaling of imaginary parts of the energy. Our results indicate that the occurrence of first excited-state $\mathcal{PT}$ transitions may be a universal feature of $\mathcal{PT}$ symmetric many-body systems. In addition, we investigate the entanglement entropy of the second-order phase transition in the real-energy regime and the $\mathcal{PT}$ transition of the full energy spectrum, and develop an approach that allows performing the finite-size scaling theory to extract the central charge even for small systems.

This paper is organized as follows. 
In Sec.\ref{sec:IsingModel}, we introduce the NHTI model in one dimension. 
In Sec.\ref{sec:entanglement}, we discuss the concept of entanglement entropy.
In Sec.\ref{sec:phasetransition}, we study the phase transitions of the NHTI model.
In Sec.\ref{sec:conclusion}, we summarize our results.

\section{Model}
\label{sec:IsingModel}
The NHTI model in one dimension is given by \cite{zhang2020ising,sun2022biorthogonal,tang2022dynamical,yang2022hidden,wang2023measurement},
\begin{equation}
	H=-\sum_{j=1}^{L}J \sigma_j^x \sigma_{j+1}^x + \sum_{j=1}^{L}h (\sigma_j^z+i\gamma \sigma_j^y),
	\label{eq:nHIsing}
\end{equation}
where $ \sigma_{j}^{\alpha} $ are three Pauli matrices along $\alpha=x, y, z$ directions at the site  $j$. 
We assume that the coupling strength $J$ and the amplitudes $h, \gamma$ of transversed fields are real numbers. 
The notation $ i=\sqrt{-1} $ is the imaginary unit. 
The periodic boundary conditions (PBCs) are imposed by $\sigma_{L+1}^{x}$ = $\sigma_{1}^{x}$ throughout the paper, with $L$ is the length of the chain.   

When $\gamma=0$, the model in Eq.(\ref{eq:nHIsing}) is the well-known Hermitian ferromagnetic transverse-field Ising model, which undergoes
a second-order phase transition between the ferromagnetic (FM) phase and the paramagnetic (PM) phase.
In the case of $\gamma \neq 0$, the Hamiltonian in Eq.(\ref{eq:nHIsing}) is non-Hermitian as $H \neq H^{\prime}$ due to the imaginary field along the $y$ direction.
However, the NHTI model has the $\mathcal{PT}$ symmetry \cite{yang2022hidden}, thus the eigenvalues are either real or complex in conjugate pairs \cite{bender1998real,bender2002complex}.  
The NHTI model in Eq.(\ref{eq:nHIsing}) can be transformed into the transverse-field Ising model,
\begin{equation}
	H=-\sum_{j=1}^{L}J \tau_j^x \tau_{j+1}^x + \sum_{j=1}^{L} h\sqrt{1-\gamma^2} \tau_j^z,
	\label{eq:Ising}
\end{equation}
through a similarity transformation \cite{zhang2020ising,yang2022hidden},
\begin{align}
\sigma_{j}^{z}= S_{j}^{-1} \tau_j^{z} S_{j},
\end{align}
on each site $j$, with $S_{j} =e^{\frac{\beta}{2}\sigma_{j}^{x}}$ and $\beta=\frac{1}{2}\ln(\frac{1+\gamma}{1-\gamma})$ denoting the transformation matrix and the non-Hermiticity.
Consequently, when $\gamma<1$, the system has real-valued energies and undergoes an Ising transition at \cite{zhang2020ising,sun2022biorthogonal,tang2022dynamical,yang2022hidden}, 
\begin{align}
\gamma_{c1} = \sqrt{1-(J/h)^{2}},
\end{align}
analogous to the Hermitian Ising model [c.f. Fig.\ref{fig:PhaseDiagram}]. 
As a simple non-Hermitian spin model, the NHTI model in Eq.(\ref{eq:nHIsing}) has been studied as a benchmark 
for topological degeneracy \cite{zhang2020ising}, biorthogonal fidelity susceptibility \cite{sun2022biorthogonal}, biorthogonal Loschmidt echo \cite{tang2022dynamical} and phase transitions without gap closing \cite{yang2022hidden} in the regime of real eigenenergies.
In the following, we will instead investigate quantum entanglement, and focus mainly on phase transitions in the complex energy regime in PBCs.

\begin{figure}[t]
	\includegraphics[width=8.7cm]{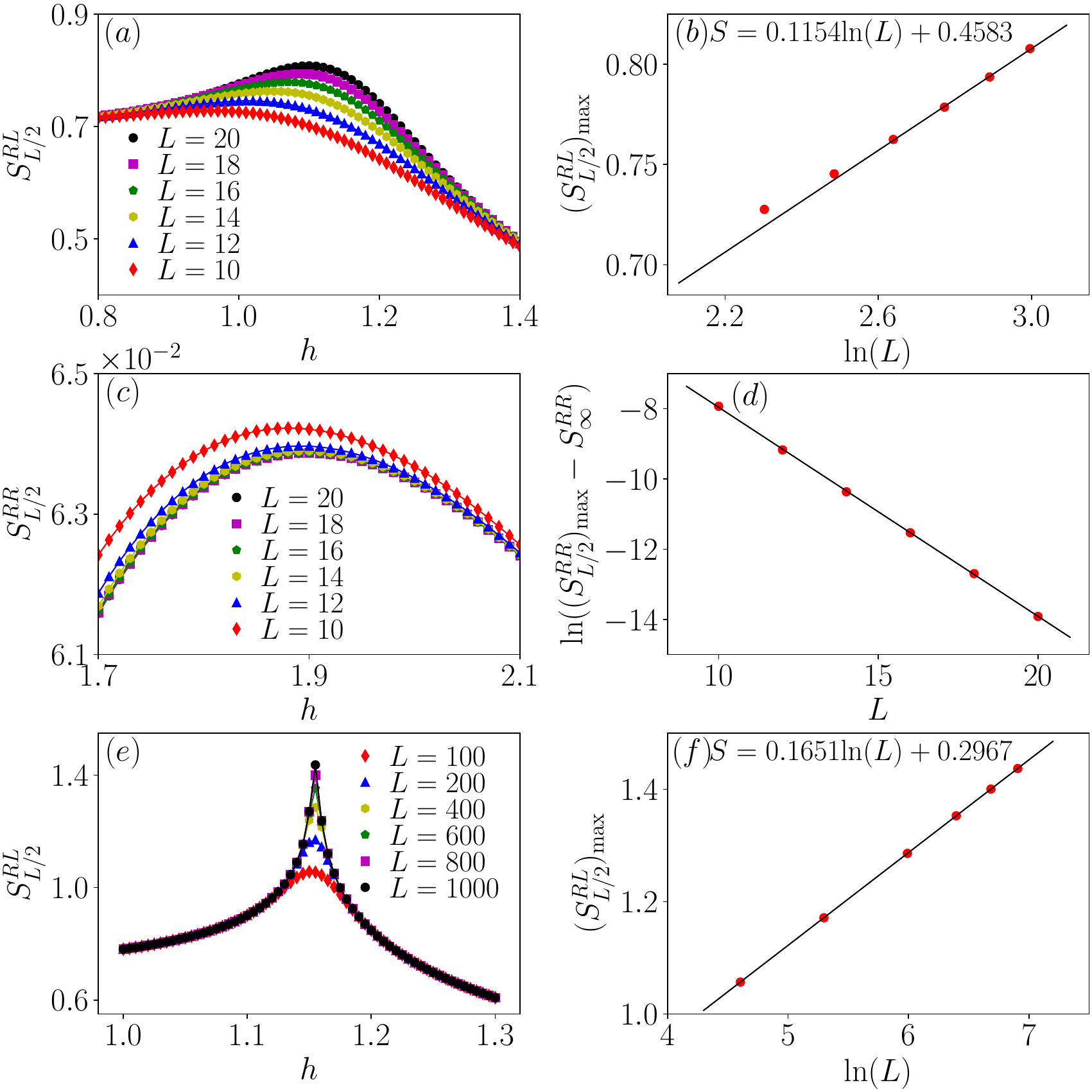}
	\caption{Entanglement entropy of ground states in the $\mathcal{PT}$ regime ($\gamma=0.5$ and $J=1$). 
	(a)(c) The biorthogonal and self-normal entanglement entropy with respect to $h$ for systems from $ L = 10 $ to $ L = 20 $. 
	(b)(d) The finite-size scaling of the biorthogonal and self-normal entanglement entropy at their peaks shown in (a) and (c), 
	which are fitted by using Eq.(\ref{equ:SScale}) and $S^{RR}_{L/2}=S^{RR}_{\infty} + a e^{-bL}$ respectively. Here, $S^{RR}_{\infty}$, $a$, $b$ are fitting parameters.
	The central charge is identified as $c=0.346$ from the biorthogonal entanglement entropy.
	(e)(f) The finite-size scaling of biorthogonal entanglement entropy for large systems up to $L=1000$, where the central charge is determined to be $c=0.495$.} 
	\label{fig:EE}
\end{figure}

\section{Entanglement}
\label{sec:entanglement}
If a system is divided into two parts, denoted as $A$ and $B$, then the von Neumann entanglement entropy $S$ of the system is defined as \cite{amico2008entanglement,eisert2010colloquium},
\begin{equation}
	S=-\text{Tr}[\rho_A \ln \rho_A]=-\text{Tr}[\rho_B \ln \rho_B],
	\label{equ:S}
\end{equation}
where the reduced density matrices $\rho_A=\text{Tr}_B(\rho)$ and $\rho_B=\text{Tr}_A(\rho)$. 
It is known that the entanglement entropy of a one-dimensional system with length $L$ near a critical point scales as \cite{calabrese2004entanglement},
\begin{equation}
	S \propto \frac{c}{3} \ln L.
	\label{equ:SScale}
\end{equation}
under PBCs, where $c$ represents the central charge. 
We will show that this definition of the entanglement entropy holds for non-Hermitian systems as well. 
However, the eigenstates $\rrangle{\phi_R}$ and $\rrangle{\phi_L}$ of $H$ and $H^{\dagger}$ are usually different as $H \neq H^{\dagger}$ in non-Hermitian systems. This indicates that two types of density matrices can be defined.

First, let us introduce the self-normal density matrix $\rho^{RR}$ using only the right eigenstates $\rrangle{\phi_R}$, which is given by,
\begin{equation}
	\rho^{RR}=\frac{\rrangle{\phi_R}\llangle{\phi_R}}{\text{Tr}(\rrangle{\phi_R}\llangle{\phi_R})}.
	\label{eq:SRR}
\end{equation}
The definition of $\rho^{RR}$ is the same as in Hermitian systems, which describes the physics originating from right eigenvectors. 
It should be noted that right eigenvectors are not all orthogonal in non-Hermitian systems, which in principle can have certain effects. 
In the framework of biorthogonal quantum mechanics \cite{brody2013biorthogonal}, one can also introduce the biorthogonal density matrix $\rho^{RL}$ defined as,
\begin{equation}
	\rho^{RL}=\frac{\rrangle{\phi_R}\llangle{\phi_L}}{\text{Tr}(\rrangle{\phi_R}\llangle{\phi_L})},
	\label{eq:SRL}
\end{equation}
where $\llangle{\phi_L} = (\rrangle{\phi_L})^{\dagger}$.
For certain phase transitions or phases, it is argued that both the biorthogonal entanglement entropy and the self-normal entanglement entropy defined respectively by $\rho^{RL}$ and $\rho^{RR}$ can work in the real-energy regime \cite{lu2023many}. In the next section, we will demonstrate that only biorthogonal entanglement entropy can characterize the phase transitions under study.

\begin{figure}[tb]
	\centering
	\includegraphics[width=8.7cm]{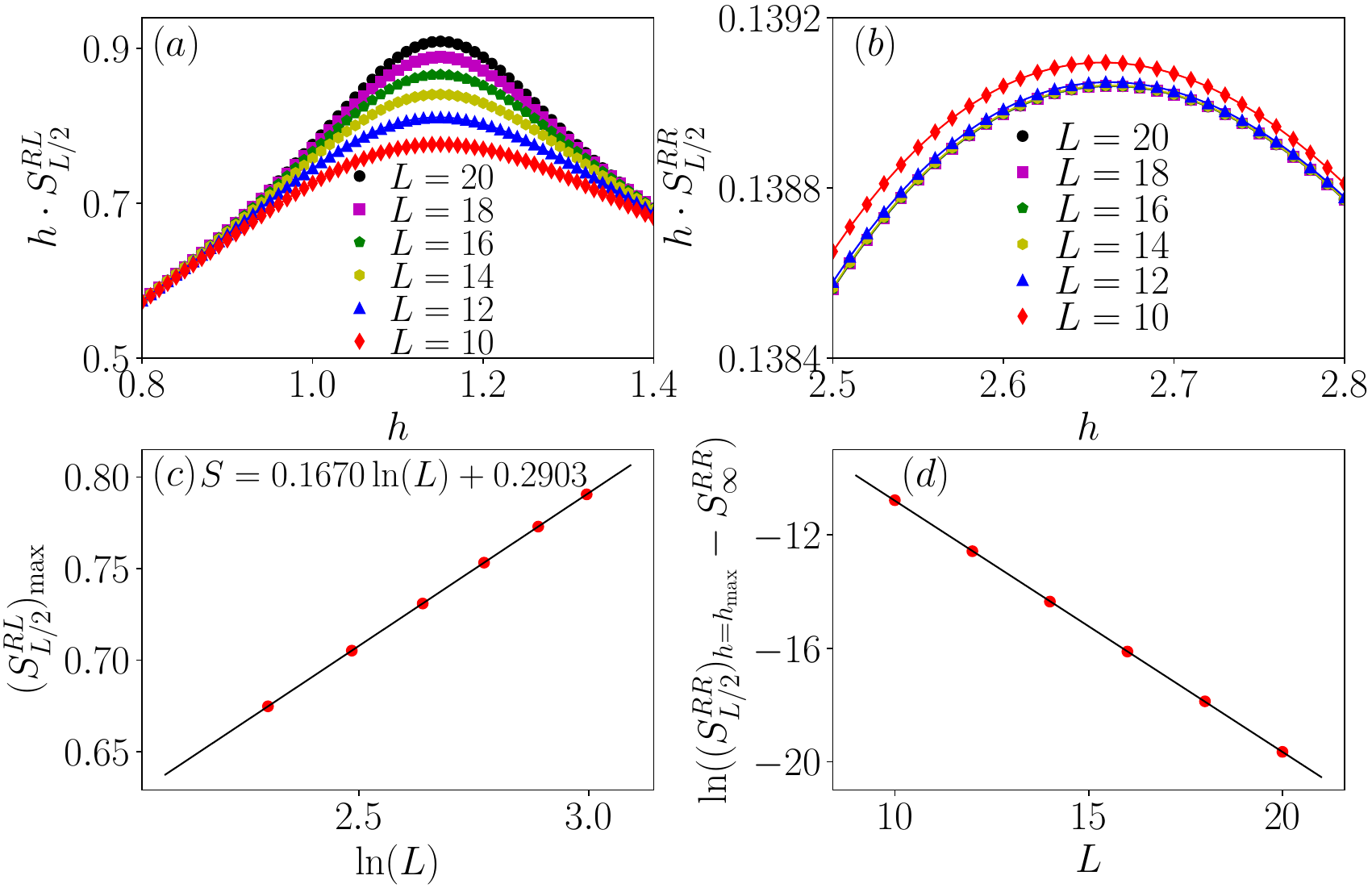}
	\caption{(a) Modified witness of the entanglement entropy of ground states in the $\mathcal{PT}$ regime ($\gamma=0.5$ and $J=1$). 
	(a)(b) The entanglement entropy $S_{L/2}^{RL}$ and $S_{L/2}^{RR}$ multiplying the factor $h$ with respect to $h$ for systems from $ L = 10 $ to $ L = 20 $. 
	(c)(d) The finite-size scaling of $S_{L/2}^{RL}$ and $S_{L/2}^{RR}$ at {\it new peaks} as shown in (a) and (b), 
	which are fitted by using Eq.(\ref{equ:SScale}) and $S^{RR}_{L/2}=S^{RR}_{\infty} + a e^{-bL}$ respectively. Here, $S^{RR}_{\infty}$, $a$, $b$ are fitting parameters.
	The central charge is derived as $c=0.501$ from $S_{L/2}^{RL}$.}
	\label{fig:EE2}
\end{figure}

\section{Phase transitions}
\label{sec:phasetransition}
\subsection{Phase transition in real-energy regime}
In order to verify whether these two entanglement entropies can characterize the phase transitions in the real-energy regime, we calculate the half-chain entanglement entropy $S_{L/2}^{RL}$ and $S_{L/2}^{RR}$ of the ground state using Eq.(\ref{eq:SRR}) and Eq.(\ref{eq:SRL}) by exact diagonalization.
Numerical results of $S_{L/2}^{RL}$ and $S_{L/2}^{RR}$ as a function of $h$ for $L = 10$ to $L = 20$ are presented in Fig.\ref{fig:EE},
where we find that the peaks of $S_{L/2}^{RL}$ obey the logarithmic scaling law [c.f. Fig.\ref{fig:EE}(a) and (b)] predicted in conformal field theory, 
while the peaks of $S_{L/2}^{RR}$ exhibit exponential decay [c.f. Fig.\ref{fig:EE} (c) and (d)].
However, as the lattice size is too small, the central charge we obtained $c=0.346$ is inconsistent with the analytical solution $c=0.5$. 
To achieve a more precise determination of the central charge, we compute $S_{L/2}^{RL}$ by mapping the spin model in Eq.(\ref{eq:Ising}) to the free fermionic Kitaev chain.
The central charge $c=0.495$ [c.f. Fig.\ref{fig:EE}(e) and (f)] of this free fermionic chain is determined using correlation functions \cite{peschel2009reduced}.
However, simulations for large interacting non-Hermitian systems are nontrivial because of the non-Hermiticity, which are under development \cite{guo2022variational,shen2023construction}.
Therefore, it would be helpful to develop an approach that use exact diagonalization to investigate non-Hermitian models in small systems. 

To achieve it, we attempt to study the scaling behavior of $h \cdot S_{L/2}^{RL}$ instead of $S_{L/2}^{RL}$ as done in Ref.\cite{sun2022biorthogonal} for the second derivative of the ground-state energy [c.f. Fig.\ref{fig:EE2}]. 
We find surprisingly that $S_{L/2}^{RL}$ obeys a logarithmic scaling law [c.f. Fig.\ref{fig:EE2}(a) and (c)] perfectly with the central charge $c = 0.501$, which matches with the analytical value $c = 0.5$. 
In contrast, $S_{L/2}^{RR}$ continues to decay exponentially and converge [c.f. Fig.\ref{fig:EE2}(b) and (d)].
Consequently, in the real-energy regime of the NHTI model, the phase transition is characterized by biorthogonal entanglement entropy rather than self-normal entanglement entropy, which is different from the phase transition in the hard-core bosonic Hatano-Nelson model, where both $S_{L/2}^{RL}$ and $S_{L/2}^{RR}$ are valid \cite{lu2023many}.

\begin{figure}[t]
	\centering
	\includegraphics[width=8.7cm]{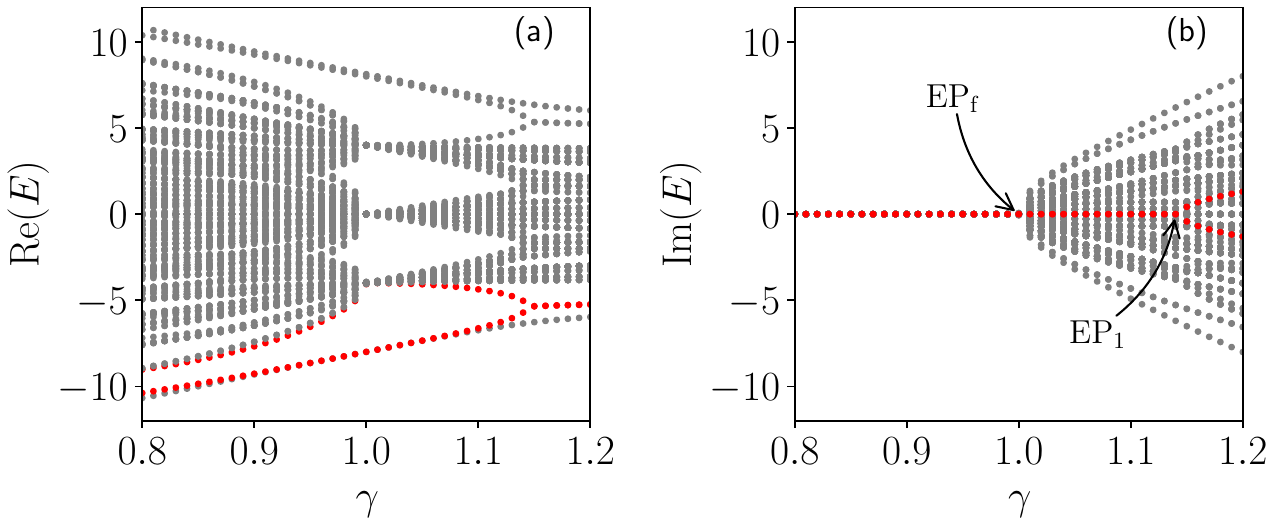}
	\caption{Energy spectrum as a function of $\gamma$ for $L=8$ at $J=1$ and $h=1.8$.
	(a) Real parts of energies, (b) Imaginary parts of energies. 
	The red dotted symbols denotes the energies of the first and second excited states. 
	$\mathrm{EP_{f}}$ denotes the full $\mathcal{PT}$ transition, and $\mathrm{EP_{1}}$ denotes the first excited-state $\mathcal{PT}$ transition.}
	\label{fig:firststateEP}
\end{figure}

\subsection {Full $\mathcal{PT}$ transition}
In a non-Hermitian system, even if the $\mathcal{PT}$ symmetry exists, real energy spectra cannot be always ensured as the $\mathcal{PT}$ symmetry can be spontaneously broken \cite{bender1998real,bender2002complex}. 
Thus, a $\mathcal{PT}$ transition with an exceptional point as its critical point can in principle occur between a $\mathcal{PT}$ symmetric phase with a real energy spectrum and a $\mathcal{PT}$ broken phase with a complex energy spectrum.
This kind of $\mathcal{PT}$ transition that characterizes the change of the full many-body energy spectrum is dubbed as {\it full $\mathcal{PT}$ transition}.
In this section we will study the full $\mathcal{PT}$ transition by looking at the energy spectrum and the biorthogonal entanglement entropy $S_{L/2}^{RL}$.

The full $\mathcal{PT}$ transition of the NHTI model is expected to occur at
\begin{equation}
\gamma_{c2} = \pm1,
\end{equation}
as the coupling coefficient of the transverse field $\tau_{j}^{z}$ in Eq.(\ref{eq:Ising}) becomes complex when $\gamma > 1$. To confirm this, we take $L = 8$ as an example to calculate the total eigenenergies of the system. 
As shown in Fig.\ref{fig:firststateEP}, we find that all energy levels are real when $0.8<\gamma<1$, while complex energy levels appear when $\gamma>1$ except for $h = 0$ which corresponds to the Hermitian Ising model.
The complete phase diagram based on the maximum imaginary energy of the system is shown in Fig.\ref{fig:PhaseDiagram}.
As expected, the eigenenergies of the system are entirely real for $|\gamma| < 1$ but become complex when $|\gamma| > 1$ independent of $h$. 
This clearly indicates that $\mathcal{PT}$ transitions occur at $|\gamma| = \pm 1$.

\begin{figure}[tb]
	\centering
	\includegraphics[width=8.7cm]{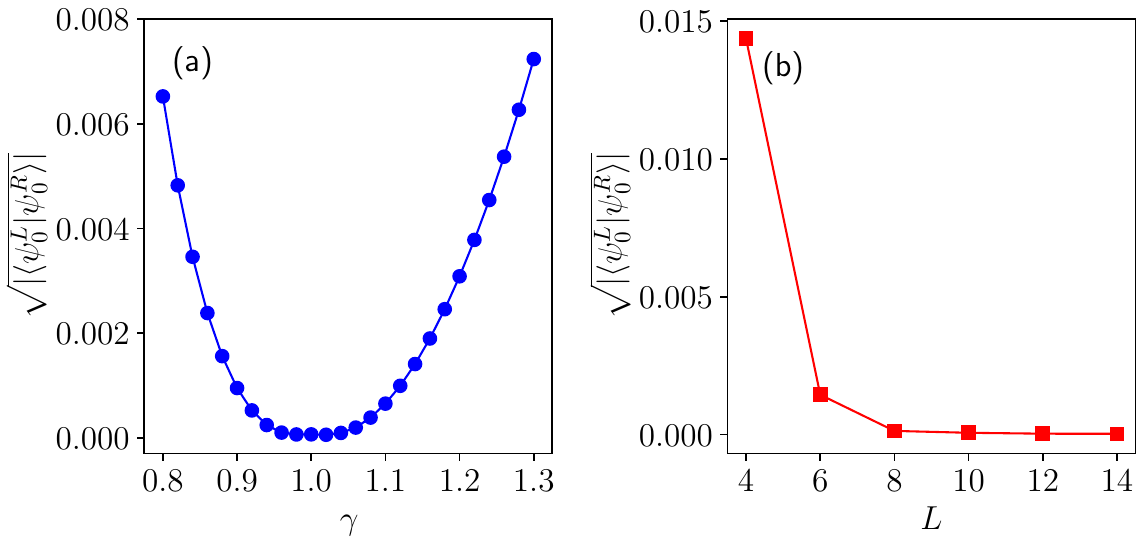}
	\caption{The biorthogonal normalization of ground states with $J=1$ and $h=1$.
	(a) The normalization factor of biorthogonal ground states as a function of $\gamma$ with $L=10$. 
	(b) The normalization factor of biorthogonal ground states as a function of $L$ near the exceptional point $\gamma=0.98$.}
	\label{fig:overlap}
\end{figure}

To investigate the properties of exceptional points, we calculate $S_{L/2}^{RL}$ near the exceptional point ($\gamma = 0.98$, $h = 1$) in the $\mathcal{PT}$ symmetric phase \cite{chang2020entanglement,guo2021entanglement,chen2022quantum,sanno2022engineering,tu2022renyi,agarwal2023recognizing}. 
However, we observe that $S_{L/2}^{RL}$ exhibits large and meaningless values.
Further investigation revealed that this anomaly stems from the ground state's degeneracy, which results in the orthogonality between the left and right eigenvectors of the ground states.
We illustrate the normalization factors of the biorthogonal eigenvectors of the ground states in Fig. \ref{fig:overlap} and observe that the normalization coefficients of biorthogonal eigenvectors near the exceptional point are small and decrease rapidly as $L$ increases.
This suggests that the computation of $S_{L/2}^{RL}$ may fail due to the ineffective determination of biorthogonal eigenvectors.

\subsection {First excited-state $\mathcal{PT}$ transition}
Phase transitions in the regime of real energies can be understood by transforming non-Hermitian Hamiltonians to their Hermitian counterparts via similarity transformations \cite{zhang2020ising,yang2022hidden,xi2021classification}.
The universality class of the phase transition under the biorthogonal basis of the NHTI model should be identical with that in the Hermitian transverse-field Ising model. 
The imaginary field in the NHTI model shifts the transition position through the real transverse field \cite{zhang2020ising}.
A fundamental question is whether a phase transition can occur in the regime of imaginary energies.
Naively speaking, this argument makes no sense, since non-Hermitian systems will amplify or attenuate if the system has complex eigenvalues.
However, we argue that it remains useful to explore the structures of energy spectra because they are related to many non-Hermitian physics, such as the non-hermitian skin effect \cite{lee2016anomalous,yao2018edge,kunst2018biorthogonal,xiong2018does,
gong2018topological,alvarez2018non,yokomizo2019non,okuma2020topological,zhang2020correspondence,yang2020non,
wang2020defective,jiang2020topological,weidemann2020topological,xiao2020non,borgnia2020non},
exceptional points \cite{heiss2012physics,kozii2024non,hodaei2017enhanced,zhou2018observation,miri2019exceptional,park2019observation,yang2019non,
ozdemir2019parity,dora2019kibble,jin2020hybrid,xiao2021observation,chen2022asymmetric} 
and non-Bloch dynamics \cite{li2023non}.

To reveal the physics of the NHTI model in the regime with imaginary energies, we rewrite the Hamiltonian in Eq.(\ref{eq:Ising}) as
\begin{equation}
	H=-\sum_{j=1}^{L} J \tau_j^x \tau_{j+1}^x + \sum_{j=1}^{L} ig \tau_j^z,
	\label{eq:model3}
\end{equation}
in the case of $\gamma > 1$, where $g = h\sqrt{\gamma^2 -1}$.
We arrive at a ferromagnetic Ising model with a purely transverse field along the $z$ direction\cite{lee2014heralded,zeng2016non,biella2021many,turkeshi2023entanglement,turkeshi2021measurement,turkeshi2022entanglement}, which can be transformed into a fermionic Kitaev chain with an imaginary potential \cite{lee2014heralded,zeng2016non},
\begin{equation}
H =-\sum_{j=1}^{L} J (c_{j}^{\dagger} c_{j+1} + c_{j}^{\dagger} c_{j+1}^{\dagger} +  \text{H.c.}) - \sum_{j=1}^{L} ig(2c_{j}^{\dagger} c_{j} -1),
\label{eq:Kitaev}
\end{equation}
by the Jordan–Wigner transformation.
The Hamiltonian in Eq.(\ref{eq:Kitaev}) can be rewritten in the Bogoliubov-de-Gennes form as,
\begin{equation}
H = \sum_{k>0} \psi_{k}^{\dagger} H_{k} \psi_{k},
\end{equation}
using the Fourier transformation in momentum space under PBCs, where
\begin{equation}
H_{k} = (-2J \cos k - i2g)\sigma^z + (2J \sin k)\sigma^y,
\label{eq:BdG}
\end{equation}
and $\psi_{k} = (c_k, c_{-k}^{\dagger})^{T}$.
The energy spectrum is given by
\begin{equation}
E_{k} = 2\sqrt{(-J \cos k - ig)^2 + (J \sin k)^2}.
\label{eq:spectrum}
\end{equation}
This spectrum has zero values in $k=\pi/2$ in contrast to $k=0$ for the normal Kitaev model with a real potential.
The energy gap closes at,
\begin{equation}
\gamma_{c3} = \pm \sqrt{1 + (J/h)^2}.
\end{equation}
Hence, the system undergoes a phase transition \cite{zeng2016non} from the FM phase ($|\gamma| < |\gamma_{c3}|$) to the PM phase ($|\gamma| > |\gamma_{c3}|$) in the spin language.
It seems that this transition is the same with the traditional Ising transition in the real regime ($\gamma < 1$).
However, we argue that this is a phase transition at which a many-body $\mathcal{PT}$ transition occurs between the first and the second excited states.
Since the single-particle energy spectrum $E_k = \sqrt{J^2- h^2(\gamma^2 -1)}$ has a $\mathcal{PT}$ transition \cite{biella2021many} at $k=\pi/2$, the eigenvalues of the first and the second many-body excited states would appear in complex conjugate pairs in the broken regime if single-particle eigenstates are labelled and analyzed according to the real parts of their energies via the Aufbau principle \cite{sun2023aufbau}. 
We note that this transition is characterized by two features: (1) the ground-state phase transition from the FM phase to the PM phase, and (2) the $\mathcal{PT}$ transition between the first and second excited states.

To verify our analytical argument, we demonstrate first excited-state $\mathcal{PT}$ transitions in Fig.\ref{fig:PhaseDiagram} with $L=8$ at $J=1$, which is obtained from many-body energy spectra as shown in Fig.\ref{fig:firststateEP}.
It can be clearly seen that the system undergoes a $\mathcal{PT}$ transition (not located at $\gamma_{c2}=\pm 1$) between the first and the second excited states.
In the case of $\gamma < \gamma_{c3}$, the first and the second excited states are separated, but when $\gamma > \gamma_{c3}$, the first and the second excited states become degenerate.
This $\mathcal{PT}$ transition is validated for various lattice sizes, indicating that it is a robust phase transition.
In addition, we find that the critical points $\gamma_{c3}$ (the positions of exceptional points) are independent of the lattice size $L$, which is different with the case discussed in Ref.\cite{zhang2022symmetry}.
Interestingly, we find that the exceptional points of the first excite-state $\mathcal{PT}$ transition and the full $\mathcal{PT}$ transition are second-order from the scaling of imaginary parts of the energies near the critical points $\gamma_{c3}$ and $\gamma_{c2}$, respectively. 
In conclusion, this phase transition, dubbed as {\it the first excited-state $\mathcal{PT}$ transition} before, is a universal and unique feature of $\mathcal{PT}$ symmetric systems.
We note that spectral transitions and the properties of the steady state with the largest imaginary part have been discussed in Ref.\cite{lee2014heralded}. Our results contribute to a comprehensive understanding of these spectral transitions, particularly focusing on the properties of the ground state and low excited states.

\section {Conclusion}
\label{sec:conclusion}
In summary, we study the biorthogonal and self-normal entanglement entropy for one-dimensional NHTI model. Through comparison, we find that the biorthogonal entanglement entropy is more suitable for characterizing the ground state phase transition than the self-normal entanglement entropy in the $\mathcal{PT}$-symmetric regime. The biorthogonal entanglement entropy exhibits logarithmic scaling behavior with the central charge $c=0.5$, while the self-normal entanglement entropy decays exponentially and converges.

Furthermore, we investigate the full $\mathcal{PT}$ phase transition. We show that the system is the $\mathcal{PT}$-symmetric phase when $|\gamma| < 1$, 
and the $\mathcal{PT}$-broken phase when $|\gamma| > 1$. Meanwhile, we study the ground-state biorthogonal entanglement entropy near the exceptional points. 
It is found that due to the degeneracy of the ground state, the entanglement entropy near the $\mathcal{PT}$ transition may be difficult to calculate.

In addition, we explore a many-body phase transition that is neither a ground-state phase transition nor a $\mathcal{PT}$ transition of the full spectrum in the $\mathcal{PT}$-broken regime.
This transition is characterized by a second-order exceptional point between the first and the second excited states,
where the critical point is independent of the system size.
Therefore, it would be interesting in the future to search for the first excited-state $\mathcal{PT}$ transition in high dimensions and to understand the universality of the transition.

\begin{acknowledgments}
We would like to thank W.-L. You and T.-C Yi for useful discussions.
G.S. is appreciative of support from the NSFC under the Grants No. 11704186,
"the Fundamental Research Funds for the Central Universities, NO. NS2023055" and 
the High Performance Computing Platform of Nanjing University of Aeronautics and Astronautics.
S.-P. K is appreciative of support from the NSFC under the Grant Nos. 11974053 and 12174030, and the National Key R \& D Program of China under Grant No. 2023YFA1406704. 
X.D. is appreciative of support from the BMBF under the Grant 13N15689.
\end{acknowledgments}

\bibliography{IsingEP}

\begin{thebibliography}{76}%
\makeatletter
\providecommand \@ifxundefined [1]{%
 \@ifx{#1\undefined}
}%
\providecommand \@ifnum [1]{%
 \ifnum #1\expandafter \@firstoftwo
 \else \expandafter \@secondoftwo
 \fi
}%
\providecommand \@ifx [1]{%
 \ifx #1\expandafter \@firstoftwo
 \else \expandafter \@secondoftwo
 \fi
}%
\providecommand \natexlab [1]{#1}%
\providecommand \enquote  [1]{``#1''}%
\providecommand \bibnamefont  [1]{#1}%
\providecommand \bibfnamefont [1]{#1}%
\providecommand \citenamefont [1]{#1}%
\providecommand \href@noop [0]{\@secondoftwo}%
\providecommand \href [0]{\begingroup \@sanitize@url \@href}%
\providecommand \@href[1]{\@@startlink{#1}\@@href}%
\providecommand \@@href[1]{\endgroup#1\@@endlink}%
\providecommand \@sanitize@url [0]{\catcode `\\12\catcode `\$12\catcode
  `\&12\catcode `\#12\catcode `\^12\catcode `\_12\catcode `\%12\relax}%
\providecommand \@@startlink[1]{}%
\providecommand \@@endlink[0]{}%
\providecommand \url  [0]{\begingroup\@sanitize@url \@url }%
\providecommand \@url [1]{\endgroup\@href {#1}{\urlprefix }}%
\providecommand \urlprefix  [0]{URL }%
\providecommand \Eprint [0]{\href }%
\providecommand \doibase [0]{https://doi.org/}%
\providecommand \selectlanguage [0]{\@gobble}%
\providecommand \bibinfo  [0]{\@secondoftwo}%
\providecommand \bibfield  [0]{\@secondoftwo}%
\providecommand \translation [1]{[#1]}%
\providecommand \BibitemOpen [0]{}%
\providecommand \bibitemStop [0]{}%
\providecommand \bibitemNoStop [0]{.\EOS\space}%
\providecommand \EOS [0]{\spacefactor3000\relax}%
\providecommand \BibitemShut  [1]{\csname bibitem#1\endcsname}%
\let\auto@bib@innerbib\@empty
\bibitem [{\citenamefont {Sachdev}(1999)}]{Sachdev1999}%
  \BibitemOpen
  \bibfield  {author} {\bibinfo {author} {\bibfnamefont {S.}~\bibnamefont
  {Sachdev}},\ }\href@noop {} {\emph {\bibinfo {title} {Quantum phase
  transitions}}}\ (\bibinfo  {publisher} {Cambridge University Press,
  Cambridge},\ \bibinfo {year} {1999})\BibitemShut {NoStop}%
\bibitem [{\citenamefont {Ginzburg}\ and\ \citenamefont
  {Landau}(2009)}]{ginzburg2009theory}%
  \BibitemOpen
  \bibfield  {author} {\bibinfo {author} {\bibfnamefont {V.~L.}\ \bibnamefont
  {Ginzburg}}\ and\ \bibinfo {author} {\bibfnamefont {L.~D.}\ \bibnamefont
  {Landau}},\ }\href@noop {} {\emph {\bibinfo {title} {On the theory of
  superconductivity}}}\ (\bibinfo  {publisher} {Springer},\ \bibinfo {year}
  {2009})\BibitemShut {NoStop}%
\bibitem [{\citenamefont {Tsuei}\ and\ \citenamefont
  {Kirtley}(2000)}]{tsuei2000pairing}%
  \BibitemOpen
  \bibfield  {author} {\bibinfo {author} {\bibfnamefont {C.}~\bibnamefont
  {Tsuei}}\ and\ \bibinfo {author} {\bibfnamefont {J.}~\bibnamefont
  {Kirtley}},\ }\bibfield  {title} {\bibinfo {title} {Pairing symmetry in
  cuprate superconductors},\ }\href
  {https://doi.org/https://doi.org/10.1103/RevModPhys.72.969} {\bibfield
  {journal} {\bibinfo  {journal} {Reviews of Modern Physics}\ }\textbf
  {\bibinfo {volume} {72}},\ \bibinfo {pages} {969} (\bibinfo {year}
  {2000})}\BibitemShut {NoStop}%
\bibitem [{\citenamefont {Wilson}\ and\ \citenamefont
  {Kogut}(1974)}]{Wilson1974}%
  \BibitemOpen
  \bibfield  {author} {\bibinfo {author} {\bibfnamefont {K.~G.}\ \bibnamefont
  {Wilson}}\ and\ \bibinfo {author} {\bibfnamefont {J.}~\bibnamefont {Kogut}},\
  }\bibfield  {title} {\bibinfo {title} {The renormalization group and the
  $\epsilon$ expansion},\ }\href {https://doi.org/10.1016/0370-1573(74)90023-4}
  {\bibfield  {journal} {\bibinfo  {journal} {Physics Reports}\ }\textbf
  {\bibinfo {volume} {12}},\ \bibinfo {pages} {75} (\bibinfo {year}
  {1974})}\BibitemShut {NoStop}%
\bibitem [{\citenamefont {Wilson}(1975)}]{Wilson1975}%
  \BibitemOpen
  \bibfield  {author} {\bibinfo {author} {\bibfnamefont {K.~G.}\ \bibnamefont
  {Wilson}},\ }\bibfield  {title} {\bibinfo {title} {The renormalization group:
  Critical phenomena and the kondo problem},\ }\href
  {https://doi.org/10.1103/RevModPhys.47.773} {\bibfield  {journal} {\bibinfo
  {journal} {Reviews of Modern Physics}\ }\textbf {\bibinfo {volume} {47}},\
  \bibinfo {pages} {773} (\bibinfo {year} {1975})}\BibitemShut {NoStop}%
\bibitem [{\citenamefont {Fisher}\ and\ \citenamefont
  {Barber}(1972)}]{Fisher1972}%
  \BibitemOpen
  \bibfield  {author} {\bibinfo {author} {\bibfnamefont {M.~E.}\ \bibnamefont
  {Fisher}}\ and\ \bibinfo {author} {\bibfnamefont {M.~N.}\ \bibnamefont
  {Barber}},\ }\bibfield  {title} {\bibinfo {title} {Scaling theory for
  finite-size effects in the critical region},\ }\href
  {https://doi.org/10.1103/PhysRevLett.28.1516} {\bibfield  {journal} {\bibinfo
   {journal} {Physical Review Letters}\ }\textbf {\bibinfo {volume} {28}},\
  \bibinfo {pages} {1516} (\bibinfo {year} {1972})}\BibitemShut {NoStop}%
\bibitem [{\citenamefont {Fisher}(1974)}]{Fisher1974}%
  \BibitemOpen
  \bibfield  {author} {\bibinfo {author} {\bibfnamefont {M.~E.}\ \bibnamefont
  {Fisher}},\ }\bibfield  {title} {\bibinfo {title} {The renormalization group
  in the theory of critical behavior},\ }\href
  {https://doi.org/10.1103/RevModPhys.46.597} {\bibfield  {journal} {\bibinfo
  {journal} {Reviews of Modern Physics}\ }\textbf {\bibinfo {volume} {46}},\
  \bibinfo {pages} {597} (\bibinfo {year} {1974})}\BibitemShut {NoStop}%
\bibitem [{\citenamefont {Heyl}\ \emph {et~al.}(2013)\citenamefont {Heyl},
  \citenamefont {Polkovnikov},\ and\ \citenamefont
  {Kehrein}}]{heyl2013dynamical}%
  \BibitemOpen
  \bibfield  {author} {\bibinfo {author} {\bibfnamefont {M.}~\bibnamefont
  {Heyl}}, \bibinfo {author} {\bibfnamefont {A.}~\bibnamefont {Polkovnikov}},\
  and\ \bibinfo {author} {\bibfnamefont {S.}~\bibnamefont {Kehrein}},\
  }\bibfield  {title} {\bibinfo {title} {Dynamical quantum phase transitions in
  the transverse-field ising model},\ }\href
  {https://doi.org/https://journals.aps.org/prl/abstract/10.1103/PhysRevLett.110.135704}
  {\bibfield  {journal} {\bibinfo  {journal} {Physical Review Letters}\
  }\textbf {\bibinfo {volume} {110}},\ \bibinfo {pages} {135704} (\bibinfo
  {year} {2013})}\BibitemShut {NoStop}%
\bibitem [{\citenamefont {Heyl}(2018)}]{heyl2018dynamical}%
  \BibitemOpen
  \bibfield  {author} {\bibinfo {author} {\bibfnamefont {M.}~\bibnamefont
  {Heyl}},\ }\bibfield  {title} {\bibinfo {title} {Dynamical quantum phase
  transitions: a review},\ }\href {https://doi.org/10.1088/1361-6633/aaaf9a}
  {\bibfield  {journal} {\bibinfo  {journal} {Reports on Progress in Physics}\
  }\textbf {\bibinfo {volume} {81}},\ \bibinfo {pages} {054001} (\bibinfo
  {year} {2018})}\BibitemShut {NoStop}%
\bibitem [{\citenamefont {Cejnar}\ \emph {et~al.}(2006)\citenamefont {Cejnar},
  \citenamefont {Macek}, \citenamefont {Heinze}, \citenamefont {Jolie},\ and\
  \citenamefont {Dobe{\v{s}}}}]{cejnar2006monodromy}%
  \BibitemOpen
  \bibfield  {author} {\bibinfo {author} {\bibfnamefont {P.}~\bibnamefont
  {Cejnar}}, \bibinfo {author} {\bibfnamefont {M.}~\bibnamefont {Macek}},
  \bibinfo {author} {\bibfnamefont {S.}~\bibnamefont {Heinze}}, \bibinfo
  {author} {\bibfnamefont {J.}~\bibnamefont {Jolie}},\ and\ \bibinfo {author}
  {\bibfnamefont {J.}~\bibnamefont {Dobe{\v{s}}}},\ }\bibfield  {title}
  {\bibinfo {title} {Monodromy and excited-state quantum phase transitions in
  integrable systems: collective vibrations of nuclei},\ }\href
  {https://doi.org/10.1088/0305-4470/39/31/L01} {\bibfield  {journal} {\bibinfo
   {journal} {Journal of Physics A: Mathematical and General}\ }\textbf
  {\bibinfo {volume} {39}},\ \bibinfo {pages} {L515} (\bibinfo {year}
  {2006})}\BibitemShut {NoStop}%
\bibitem [{\citenamefont {Cejnar}\ \emph {et~al.}(2007)\citenamefont {Cejnar},
  \citenamefont {Heinze},\ and\ \citenamefont {Macek}}]{cejnar2007coulomb}%
  \BibitemOpen
  \bibfield  {author} {\bibinfo {author} {\bibfnamefont {P.}~\bibnamefont
  {Cejnar}}, \bibinfo {author} {\bibfnamefont {S.}~\bibnamefont {Heinze}},\
  and\ \bibinfo {author} {\bibfnamefont {M.}~\bibnamefont {Macek}},\ }\bibfield
   {title} {\bibinfo {title} {Coulomb analogy for non-hermitian degeneracies
  near quantum phase transitions},\ }\href
  {https://doi.org/https://doi.org/10.1103/PhysRevLett.99.100601} {\bibfield
  {journal} {\bibinfo  {journal} {Physical Review Letters}\ }\textbf {\bibinfo
  {volume} {99}},\ \bibinfo {pages} {100601} (\bibinfo {year}
  {2007})}\BibitemShut {NoStop}%
\bibitem [{\citenamefont {Caprio}\ \emph {et~al.}(2008)\citenamefont {Caprio},
  \citenamefont {Cejnar},\ and\ \citenamefont {Iachello}}]{caprio2008excited}%
  \BibitemOpen
  \bibfield  {author} {\bibinfo {author} {\bibfnamefont {M.}~\bibnamefont
  {Caprio}}, \bibinfo {author} {\bibfnamefont {P.}~\bibnamefont {Cejnar}},\
  and\ \bibinfo {author} {\bibfnamefont {F.}~\bibnamefont {Iachello}},\
  }\bibfield  {title} {\bibinfo {title} {Excited state quantum phase
  transitions in many-body systems},\ }\href
  {https://doi.org/https://doi.org/10.1016/j.aop.2007.06.011} {\bibfield
  {journal} {\bibinfo  {journal} {Annals of Physics}\ }\textbf {\bibinfo
  {volume} {323}},\ \bibinfo {pages} {1106} (\bibinfo {year}
  {2008})}\BibitemShut {NoStop}%
\bibitem [{\citenamefont {Puebla}\ \emph {et~al.}(2016)\citenamefont {Puebla},
  \citenamefont {Hwang},\ and\ \citenamefont {Plenio}}]{puebla2016excited}%
  \BibitemOpen
  \bibfield  {author} {\bibinfo {author} {\bibfnamefont {R.}~\bibnamefont
  {Puebla}}, \bibinfo {author} {\bibfnamefont {M.-J.}\ \bibnamefont {Hwang}},\
  and\ \bibinfo {author} {\bibfnamefont {M.~B.}\ \bibnamefont {Plenio}},\
  }\bibfield  {title} {\bibinfo {title} {Excited-state quantum phase transition
  in the rabi model},\ }\href
  {https://doi.org/https://doi.org/10.1103/PhysRevA.94.023835} {\bibfield
  {journal} {\bibinfo  {journal} {Physical Review A}\ }\textbf {\bibinfo
  {volume} {94}},\ \bibinfo {pages} {023835} (\bibinfo {year}
  {2016})}\BibitemShut {NoStop}%
\bibitem [{\citenamefont {Feldmann}\ \emph {et~al.}(2021)\citenamefont
  {Feldmann}, \citenamefont {Klempt}, \citenamefont {Smerzi}, \citenamefont
  {Santos},\ and\ \citenamefont {Gessner}}]{feldmann2021interferometric}%
  \BibitemOpen
  \bibfield  {author} {\bibinfo {author} {\bibfnamefont {P.}~\bibnamefont
  {Feldmann}}, \bibinfo {author} {\bibfnamefont {C.}~\bibnamefont {Klempt}},
  \bibinfo {author} {\bibfnamefont {A.}~\bibnamefont {Smerzi}}, \bibinfo
  {author} {\bibfnamefont {L.}~\bibnamefont {Santos}},\ and\ \bibinfo {author}
  {\bibfnamefont {M.}~\bibnamefont {Gessner}},\ }\bibfield  {title} {\bibinfo
  {title} {Interferometric order parameter for excited-state quantum phase
  transitions in bose-einstein condensates},\ }\href
  {https://doi.org/https://doi.org/10.1103/PhysRevLett.126.230602} {\bibfield
  {journal} {\bibinfo  {journal} {Physical Review Letters}\ }\textbf {\bibinfo
  {volume} {126}},\ \bibinfo {pages} {230602} (\bibinfo {year}
  {2021})}\BibitemShut {NoStop}%
\bibitem [{\citenamefont {Cejnar}\ \emph {et~al.}(2021)\citenamefont {Cejnar},
  \citenamefont {Str{\'a}nsk{\`y}}, \citenamefont {Macek},\ and\ \citenamefont
  {Kloc}}]{cejnar2021excited}%
  \BibitemOpen
  \bibfield  {author} {\bibinfo {author} {\bibfnamefont {P.}~\bibnamefont
  {Cejnar}}, \bibinfo {author} {\bibfnamefont {P.}~\bibnamefont
  {Str{\'a}nsk{\`y}}}, \bibinfo {author} {\bibfnamefont {M.}~\bibnamefont
  {Macek}},\ and\ \bibinfo {author} {\bibfnamefont {M.}~\bibnamefont {Kloc}},\
  }\bibfield  {title} {\bibinfo {title} {Excited-state quantum phase
  transitions},\ }\href {https://doi.org/10.1088/1751-8121/abdfe8} {\bibfield
  {journal} {\bibinfo  {journal} {Journal of Physics A: Mathematical and
  Theoretical}\ }\textbf {\bibinfo {volume} {54}},\ \bibinfo {pages} {133001}
  (\bibinfo {year} {2021})}\BibitemShut {NoStop}%
\bibitem [{\citenamefont {Bender}\ and\ \citenamefont
  {Boettcher}(1998)}]{bender1998real}%
  \BibitemOpen
  \bibfield  {author} {\bibinfo {author} {\bibfnamefont {C.~M.}\ \bibnamefont
  {Bender}}\ and\ \bibinfo {author} {\bibfnamefont {S.}~\bibnamefont
  {Boettcher}},\ }\bibfield  {title} {\bibinfo {title} {Real spectra in
  non-hermitian hamiltonians having $\mathcal{PT}$ symmetry},\ }\href
  {https://doi.org/https://doi.org/10.1103/PhysRevLett.80.5243} {\bibfield
  {journal} {\bibinfo  {journal} {Physical Review Letters}\ }\textbf {\bibinfo
  {volume} {80}},\ \bibinfo {pages} {5243} (\bibinfo {year}
  {1998})}\BibitemShut {NoStop}%
\bibitem [{\citenamefont {Bender}\ \emph {et~al.}(2002)\citenamefont {Bender},
  \citenamefont {Brody},\ and\ \citenamefont {Jones}}]{bender2002complex}%
  \BibitemOpen
  \bibfield  {author} {\bibinfo {author} {\bibfnamefont {C.~M.}\ \bibnamefont
  {Bender}}, \bibinfo {author} {\bibfnamefont {D.~C.}\ \bibnamefont {Brody}},\
  and\ \bibinfo {author} {\bibfnamefont {H.~F.}\ \bibnamefont {Jones}},\
  }\bibfield  {title} {\bibinfo {title} {Complex extension of quantum
  mechanics},\ }\href
  {https://doi.org/https://doi.org/10.1103/PhysRevLett.89.270401} {\bibfield
  {journal} {\bibinfo  {journal} {Physical Review Letters}\ }\textbf {\bibinfo
  {volume} {89}},\ \bibinfo {pages} {270401} (\bibinfo {year}
  {2002})}\BibitemShut {NoStop}%
\bibitem [{\citenamefont {Bergholtz}\ \emph {et~al.}(2021)\citenamefont
  {Bergholtz}, \citenamefont {Budich},\ and\ \citenamefont
  {Kunst}}]{bergholtz2021exceptional}%
  \BibitemOpen
  \bibfield  {author} {\bibinfo {author} {\bibfnamefont {E.~J.}\ \bibnamefont
  {Bergholtz}}, \bibinfo {author} {\bibfnamefont {J.~C.}\ \bibnamefont
  {Budich}},\ and\ \bibinfo {author} {\bibfnamefont {F.~K.}\ \bibnamefont
  {Kunst}},\ }\bibfield  {title} {\bibinfo {title} {Exceptional topology of
  non-hermitian systems},\ }\href
  {https://doi.org/https://doi.org/10.1103/RevModPhys.93.015005} {\bibfield
  {journal} {\bibinfo  {journal} {Reviews of Modern Physics}\ }\textbf
  {\bibinfo {volume} {93}},\ \bibinfo {pages} {015005} (\bibinfo {year}
  {2021})}\BibitemShut {NoStop}%
\bibitem [{\citenamefont {Ashida}\ \emph {et~al.}(2021)\citenamefont {Ashida},
  \citenamefont {Gong},\ and\ \citenamefont {Ueda}}]{ashida2021non}%
  \BibitemOpen
  \bibfield  {author} {\bibinfo {author} {\bibfnamefont {Y.}~\bibnamefont
  {Ashida}}, \bibinfo {author} {\bibfnamefont {Z.}~\bibnamefont {Gong}},\ and\
  \bibinfo {author} {\bibfnamefont {M.}~\bibnamefont {Ueda}},\ }\bibfield
  {title} {\bibinfo {title} {Non-hermitian physics},\ }\href
  {https://doi.org/https://doi.org/10.1080/00018732.2021.1876991} {\bibfield
  {journal} {\bibinfo  {journal} {Advances in Physics}\ }\textbf {\bibinfo
  {volume} {69}},\ \bibinfo {pages} {249} (\bibinfo {year} {2021})}\BibitemShut
  {NoStop}%
\bibitem [{\citenamefont {Zhang}\ \emph {et~al.}(2022)\citenamefont {Zhang},
  \citenamefont {Denner}, \citenamefont {Bzdu{\v{s}}ek}, \citenamefont
  {Sentef},\ and\ \citenamefont {Neupert}}]{zhang2022symmetry}%
  \BibitemOpen
  \bibfield  {author} {\bibinfo {author} {\bibfnamefont {S.-B.}\ \bibnamefont
  {Zhang}}, \bibinfo {author} {\bibfnamefont {M.~M.}\ \bibnamefont {Denner}},
  \bibinfo {author} {\bibfnamefont {T.}~\bibnamefont {Bzdu{\v{s}}ek}}, \bibinfo
  {author} {\bibfnamefont {M.~A.}\ \bibnamefont {Sentef}},\ and\ \bibinfo
  {author} {\bibfnamefont {T.}~\bibnamefont {Neupert}},\ }\bibfield  {title}
  {\bibinfo {title} {Symmetry breaking and spectral structure of the
  interacting hatano-nelson model},\ }\href
  {https://doi.org/https://doi.org/10.1103/PhysRevB.106.L121102} {\bibfield
  {journal} {\bibinfo  {journal} {Physical Review B}\ }\textbf {\bibinfo
  {volume} {106}},\ \bibinfo {pages} {L121102} (\bibinfo {year}
  {2022})}\BibitemShut {NoStop}%
\bibitem [{\citenamefont {Lu}\ and\ \citenamefont {Sun}(2023)}]{lu2023many}%
  \BibitemOpen
  \bibfield  {author} {\bibinfo {author} {\bibfnamefont {C.-Z.}\ \bibnamefont
  {Lu}}\ and\ \bibinfo {author} {\bibfnamefont {G.}~\bibnamefont {Sun}},\
  }\bibfield  {title} {\bibinfo {title} {Many-body entanglement and spectral
  clusters in the extended hard-core bosonic hatano-nelson model},\ }\href
  {https://arxiv.org/abs/2310.07599} {\bibfield  {journal} {\bibinfo  {journal}
  {arXiv:2310.07599}\ } (\bibinfo {year} {2023})}\BibitemShut {NoStop}%
\bibitem [{\citenamefont {Yang}\ and\ \citenamefont
  {Luo}(2023)}]{yang2023first}%
  \BibitemOpen
  \bibfield  {author} {\bibinfo {author} {\bibfnamefont {Y.-T.}\ \bibnamefont
  {Yang}}\ and\ \bibinfo {author} {\bibfnamefont {H.-G.}\ \bibnamefont {Luo}},\
  }\bibfield  {title} {\bibinfo {title} {First-order excited-state quantum
  phase transition in the transverse ising model with a longitudinal field},\
  }\href {https://arxiv.org/abs/2301.02066} {\bibfield  {journal} {\bibinfo
  {journal} {arXiv:2301.02066}\ } (\bibinfo {year} {2023})}\BibitemShut
  {NoStop}%
\bibitem [{\citenamefont {Zhang}\ and\ \citenamefont
  {Song}(2020)}]{zhang2020ising}%
  \BibitemOpen
  \bibfield  {author} {\bibinfo {author} {\bibfnamefont {K.}~\bibnamefont
  {Zhang}}\ and\ \bibinfo {author} {\bibfnamefont {Z.}~\bibnamefont {Song}},\
  }\bibfield  {title} {\bibinfo {title} {Ising chain with topological
  degeneracy induced by dissipation},\ }\href
  {https://doi.org/https://doi.org/10.1103/PhysRevB.101.245152} {\bibfield
  {journal} {\bibinfo  {journal} {Physical Review B}\ }\textbf {\bibinfo
  {volume} {101}},\ \bibinfo {pages} {245152} (\bibinfo {year}
  {2020})}\BibitemShut {NoStop}%
\bibitem [{\citenamefont {Sun}\ \emph {et~al.}(2022)\citenamefont {Sun},
  \citenamefont {Tang},\ and\ \citenamefont {Kou}}]{sun2022biorthogonal}%
  \BibitemOpen
  \bibfield  {author} {\bibinfo {author} {\bibfnamefont {G.}~\bibnamefont
  {Sun}}, \bibinfo {author} {\bibfnamefont {J.-C.}\ \bibnamefont {Tang}},\ and\
  \bibinfo {author} {\bibfnamefont {S.-P.}\ \bibnamefont {Kou}},\ }\bibfield
  {title} {\bibinfo {title} {Biorthogonal quantum criticality in non-hermitian
  many-body systems},\ }\href
  {https://doi.org/https://doi.org/10.1007/s11467-021-1126-1} {\bibfield
  {journal} {\bibinfo  {journal} {Frontiers of Physics}\ }\textbf {\bibinfo
  {volume} {17}},\ \bibinfo {pages} {1} (\bibinfo {year} {2022})}\BibitemShut
  {NoStop}%
\bibitem [{\citenamefont {Tang}\ \emph {et~al.}(2022)\citenamefont {Tang},
  \citenamefont {Kou},\ and\ \citenamefont {Sun}}]{tang2022dynamical}%
  \BibitemOpen
  \bibfield  {author} {\bibinfo {author} {\bibfnamefont {J.-C.}\ \bibnamefont
  {Tang}}, \bibinfo {author} {\bibfnamefont {S.-P.}\ \bibnamefont {Kou}},\ and\
  \bibinfo {author} {\bibfnamefont {G.}~\bibnamefont {Sun}},\ }\bibfield
  {title} {\bibinfo {title} {Dynamical scaling of loschmidt echo in
  non-hermitian systems},\ }\href {https://doi.org/10.1209/0295-5075/ac53c4}
  {\bibfield  {journal} {\bibinfo  {journal} {Europhysics Letters}\ }\textbf
  {\bibinfo {volume} {137}},\ \bibinfo {pages} {40001} (\bibinfo {year}
  {2022})}\BibitemShut {NoStop}%
\bibitem [{\citenamefont {Yang}\ \emph {et~al.}(2022)\citenamefont {Yang},
  \citenamefont {Wang}, \citenamefont {Yang}, \citenamefont {Guo},
  \citenamefont {Wang}, \citenamefont {Sun},\ and\ \citenamefont
  {Kou}}]{yang2022hidden}%
  \BibitemOpen
  \bibfield  {author} {\bibinfo {author} {\bibfnamefont {F.}~\bibnamefont
  {Yang}}, \bibinfo {author} {\bibfnamefont {H.}~\bibnamefont {Wang}}, \bibinfo
  {author} {\bibfnamefont {M.-L.}\ \bibnamefont {Yang}}, \bibinfo {author}
  {\bibfnamefont {C.-X.}\ \bibnamefont {Guo}}, \bibinfo {author} {\bibfnamefont
  {X.-R.}\ \bibnamefont {Wang}}, \bibinfo {author} {\bibfnamefont {G.-Y.}\
  \bibnamefont {Sun}},\ and\ \bibinfo {author} {\bibfnamefont {S.-P.}\
  \bibnamefont {Kou}},\ }\bibfield  {title} {\bibinfo {title} {Hidden
  continuous quantum phase transition without gap closing in non-hermitian
  transverse ising model},\ }\href {https://doi.org/10.1088/1367-2630/ac652f}
  {\bibfield  {journal} {\bibinfo  {journal} {New Journal of Physics}\ }\textbf
  {\bibinfo {volume} {24}},\ \bibinfo {pages} {043046} (\bibinfo {year}
  {2022})}\BibitemShut {NoStop}%
\bibitem [{\citenamefont {Wang}\ \emph {et~al.}(2023)\citenamefont {Wang},
  \citenamefont {Zhu}, \citenamefont {Lang},\ and\ \citenamefont
  {He}}]{wang2023measurement}%
  \BibitemOpen
  \bibfield  {author} {\bibinfo {author} {\bibfnamefont {Z.}~\bibnamefont
  {Wang}}, \bibinfo {author} {\bibfnamefont {S.-L.}\ \bibnamefont {Zhu}},
  \bibinfo {author} {\bibfnamefont {L.-J.}\ \bibnamefont {Lang}},\ and\
  \bibinfo {author} {\bibfnamefont {L.}~\bibnamefont {He}},\ }\bibfield
  {title} {\bibinfo {title} {Measurement-induced integer families of critical
  dynamical scaling in quantum many-body systems},\ }\href
  {https://arxiv.org/abs/2308.06567} {\bibfield  {journal} {\bibinfo  {journal}
  {arXiv:2308.06567}\ } (\bibinfo {year} {2023})}\BibitemShut {NoStop}%
\bibitem [{\citenamefont {Amico}\ \emph {et~al.}(2008)\citenamefont {Amico},
  \citenamefont {Fazio}, \citenamefont {Osterloh},\ and\ \citenamefont
  {Vedral}}]{amico2008entanglement}%
  \BibitemOpen
  \bibfield  {author} {\bibinfo {author} {\bibfnamefont {L.}~\bibnamefont
  {Amico}}, \bibinfo {author} {\bibfnamefont {R.}~\bibnamefont {Fazio}},
  \bibinfo {author} {\bibfnamefont {A.}~\bibnamefont {Osterloh}},\ and\
  \bibinfo {author} {\bibfnamefont {V.}~\bibnamefont {Vedral}},\ }\bibfield
  {title} {\bibinfo {title} {Entanglement in many-body systems},\ }\href
  {https://doi.org/https://doi.org/10.1103/RevModPhys.80.517} {\bibfield
  {journal} {\bibinfo  {journal} {Reviews of Modern Physics}\ }\textbf
  {\bibinfo {volume} {80}},\ \bibinfo {pages} {517} (\bibinfo {year}
  {2008})}\BibitemShut {NoStop}%
\bibitem [{\citenamefont {Eisert}\ \emph {et~al.}(2010)\citenamefont {Eisert},
  \citenamefont {Cramer},\ and\ \citenamefont {Plenio}}]{eisert2010colloquium}%
  \BibitemOpen
  \bibfield  {author} {\bibinfo {author} {\bibfnamefont {J.}~\bibnamefont
  {Eisert}}, \bibinfo {author} {\bibfnamefont {M.}~\bibnamefont {Cramer}},\
  and\ \bibinfo {author} {\bibfnamefont {M.~B.}\ \bibnamefont {Plenio}},\
  }\bibfield  {title} {\bibinfo {title} {Colloquium: Area laws for the
  entanglement entropy},\ }\href
  {https://doi.org/https://doi.org/10.1103/RevModPhys.82.277} {\bibfield
  {journal} {\bibinfo  {journal} {Reviews of Modern Physics}\ }\textbf
  {\bibinfo {volume} {82}},\ \bibinfo {pages} {277} (\bibinfo {year}
  {2010})}\BibitemShut {NoStop}%
\bibitem [{\citenamefont {Calabrese}\ and\ \citenamefont
  {Cardy}(2004)}]{calabrese2004entanglement}%
  \BibitemOpen
  \bibfield  {author} {\bibinfo {author} {\bibfnamefont {P.}~\bibnamefont
  {Calabrese}}\ and\ \bibinfo {author} {\bibfnamefont {J.}~\bibnamefont
  {Cardy}},\ }\bibfield  {title} {\bibinfo {title} {Entanglement entropy and
  quantum field theory},\ }\href
  {https://doi.org/10.1088/1742-5468/2004/06/P06002} {\bibfield  {journal}
  {\bibinfo  {journal} {Journal of Statistical Mechanics: Theory and
  Experiment}\ }\textbf {\bibinfo {volume} {2004}},\ \bibinfo {pages} {P06002}
  (\bibinfo {year} {2004})}\BibitemShut {NoStop}%
\bibitem [{\citenamefont {Brody}(2013)}]{brody2013biorthogonal}%
  \BibitemOpen
  \bibfield  {author} {\bibinfo {author} {\bibfnamefont {D.~C.}\ \bibnamefont
  {Brody}},\ }\bibfield  {title} {\bibinfo {title} {Biorthogonal quantum
  mechanics},\ }\href {https://doi.org/10.1088/1751-8113/47/3/035305}
  {\bibfield  {journal} {\bibinfo  {journal} {Journal of Physics A:
  Mathematical and Theoretical}\ }\textbf {\bibinfo {volume} {47}},\ \bibinfo
  {pages} {035305} (\bibinfo {year} {2013})}\BibitemShut {NoStop}%
\bibitem [{\citenamefont {Peschel}\ and\ \citenamefont
  {Eisler}(2009)}]{peschel2009reduced}%
  \BibitemOpen
  \bibfield  {author} {\bibinfo {author} {\bibfnamefont {I.}~\bibnamefont
  {Peschel}}\ and\ \bibinfo {author} {\bibfnamefont {V.}~\bibnamefont
  {Eisler}},\ }\bibfield  {title} {\bibinfo {title} {Reduced density matrices
  and entanglement entropy in free lattice models},\ }\href
  {https://doi.org/10.1088/1751-8113/42/50/504003} {\bibfield  {journal}
  {\bibinfo  {journal} {Journal of physics a: mathematical and theoretical}\
  }\textbf {\bibinfo {volume} {42}},\ \bibinfo {pages} {504003} (\bibinfo
  {year} {2009})}\BibitemShut {NoStop}%
\bibitem [{\citenamefont {Guo}\ \emph {et~al.}(2022)\citenamefont {Guo},
  \citenamefont {Xu}, \citenamefont {Li}, \citenamefont {You},\ and\
  \citenamefont {Yang}}]{guo2022variational}%
  \BibitemOpen
  \bibfield  {author} {\bibinfo {author} {\bibfnamefont {Z.}~\bibnamefont
  {Guo}}, \bibinfo {author} {\bibfnamefont {Z.-T.}\ \bibnamefont {Xu}},
  \bibinfo {author} {\bibfnamefont {M.}~\bibnamefont {Li}}, \bibinfo {author}
  {\bibfnamefont {L.}~\bibnamefont {You}},\ and\ \bibinfo {author}
  {\bibfnamefont {S.}~\bibnamefont {Yang}},\ }\bibfield  {title} {\bibinfo
  {title} {Variational matrix product state approach for non-hermitian system
  based on a companion hermitian hamiltonian},\ }\href
  {https://arxiv.org/abs/2210.14858} {\bibfield  {journal} {\bibinfo  {journal}
  {arXiv:2210.14858}\ } (\bibinfo {year} {2022})}\BibitemShut {NoStop}%
\bibitem [{\citenamefont {Shen}\ \emph {et~al.}(2023)\citenamefont {Shen},
  \citenamefont {Guo},\ and\ \citenamefont {Yang}}]{shen2023construction}%
  \BibitemOpen
  \bibfield  {author} {\bibinfo {author} {\bibfnamefont {R.}~\bibnamefont
  {Shen}}, \bibinfo {author} {\bibfnamefont {Y.}~\bibnamefont {Guo}},\ and\
  \bibinfo {author} {\bibfnamefont {S.}~\bibnamefont {Yang}},\ }\bibfield
  {title} {\bibinfo {title} {Construction of non-hermitian parent hamiltonian
  from matrix product states},\ }\href
  {https://doi.org/https://doi.org/10.1103/PhysRevLett.130.220401} {\bibfield
  {journal} {\bibinfo  {journal} {Physical Review Letters}\ }\textbf {\bibinfo
  {volume} {130}},\ \bibinfo {pages} {220401} (\bibinfo {year}
  {2023})}\BibitemShut {NoStop}%
\bibitem [{\citenamefont {Chang}\ \emph {et~al.}(2020)\citenamefont {Chang},
  \citenamefont {You}, \citenamefont {Wen},\ and\ \citenamefont
  {Ryu}}]{chang2020entanglement}%
  \BibitemOpen
  \bibfield  {author} {\bibinfo {author} {\bibfnamefont {P.-Y.}\ \bibnamefont
  {Chang}}, \bibinfo {author} {\bibfnamefont {J.-S.}\ \bibnamefont {You}},
  \bibinfo {author} {\bibfnamefont {X.}~\bibnamefont {Wen}},\ and\ \bibinfo
  {author} {\bibfnamefont {S.}~\bibnamefont {Ryu}},\ }\bibfield  {title}
  {\bibinfo {title} {Entanglement spectrum and entropy in topological
  non-hermitian systems and nonunitary conformal field theory},\ }\href
  {https://doi.org/https://doi.org/10.1103/PhysRevResearch.2.033069} {\bibfield
   {journal} {\bibinfo  {journal} {Physical Review Research}\ }\textbf
  {\bibinfo {volume} {2}},\ \bibinfo {pages} {033069} (\bibinfo {year}
  {2020})}\BibitemShut {NoStop}%
\bibitem [{\citenamefont {Guo}\ \emph {et~al.}(2021)\citenamefont {Guo},
  \citenamefont {Yu}, \citenamefont {Huang}, \citenamefont {Yang},
  \citenamefont {Chi}, \citenamefont {Liao},\ and\ \citenamefont
  {Xiang}}]{guo2021entanglement}%
  \BibitemOpen
  \bibfield  {author} {\bibinfo {author} {\bibfnamefont {Y.-B.}\ \bibnamefont
  {Guo}}, \bibinfo {author} {\bibfnamefont {Y.-C.}\ \bibnamefont {Yu}},
  \bibinfo {author} {\bibfnamefont {R.-Z.}\ \bibnamefont {Huang}}, \bibinfo
  {author} {\bibfnamefont {L.-P.}\ \bibnamefont {Yang}}, \bibinfo {author}
  {\bibfnamefont {R.-Z.}\ \bibnamefont {Chi}}, \bibinfo {author} {\bibfnamefont
  {H.-J.}\ \bibnamefont {Liao}},\ and\ \bibinfo {author} {\bibfnamefont
  {T.}~\bibnamefont {Xiang}},\ }\bibfield  {title} {\bibinfo {title}
  {Entanglement entropy of non-hermitian free fermions},\ }\href
  {https://doi.org/10.1088/1361-648X/ac216e} {\bibfield  {journal} {\bibinfo
  {journal} {Journal of Physics: Condensed Matter}\ }\textbf {\bibinfo {volume}
  {33}},\ \bibinfo {pages} {475502} (\bibinfo {year} {2021})}\BibitemShut
  {NoStop}%
\bibitem [{\citenamefont {Chen}\ \emph
  {et~al.}(2022{\natexlab{a}})\citenamefont {Chen}, \citenamefont {Zhou},
  \citenamefont {Chen},\ and\ \citenamefont {Ye}}]{chen2022quantum}%
  \BibitemOpen
  \bibfield  {author} {\bibinfo {author} {\bibfnamefont {L.-M.}\ \bibnamefont
  {Chen}}, \bibinfo {author} {\bibfnamefont {Y.}~\bibnamefont {Zhou}}, \bibinfo
  {author} {\bibfnamefont {S.~A.}\ \bibnamefont {Chen}},\ and\ \bibinfo
  {author} {\bibfnamefont {P.}~\bibnamefont {Ye}},\ }\bibfield  {title}
  {\bibinfo {title} {Quantum entanglement of non-hermitian quasicrystals},\
  }\href {https://doi.org/https://doi.org/10.1103/PhysRevB.105.L121115}
  {\bibfield  {journal} {\bibinfo  {journal} {Physical Review B}\ }\textbf
  {\bibinfo {volume} {105}},\ \bibinfo {pages} {L121115} (\bibinfo {year}
  {2022}{\natexlab{a}})}\BibitemShut {NoStop}%
\bibitem [{\citenamefont {Sanno}\ \emph {et~al.}(2022)\citenamefont {Sanno},
  \citenamefont {Yamada}, \citenamefont {Mizushima},\ and\ \citenamefont
  {Fujimoto}}]{sanno2022engineering}%
  \BibitemOpen
  \bibfield  {author} {\bibinfo {author} {\bibfnamefont {T.}~\bibnamefont
  {Sanno}}, \bibinfo {author} {\bibfnamefont {M.~G.}\ \bibnamefont {Yamada}},
  \bibinfo {author} {\bibfnamefont {T.}~\bibnamefont {Mizushima}},\ and\
  \bibinfo {author} {\bibfnamefont {S.}~\bibnamefont {Fujimoto}},\ }\bibfield
  {title} {\bibinfo {title} {Engineering yang-lee anyons via majorana bound
  states},\ }\href
  {https://doi.org/https://doi.org/10.1103/PhysRevB.106.174517} {\bibfield
  {journal} {\bibinfo  {journal} {Physical Review B}\ }\textbf {\bibinfo
  {volume} {106}},\ \bibinfo {pages} {174517} (\bibinfo {year}
  {2022})}\BibitemShut {NoStop}%
\bibitem [{\citenamefont {Tu}\ \emph {et~al.}(2022)\citenamefont {Tu},
  \citenamefont {Tzeng},\ and\ \citenamefont {Chang}}]{tu2022renyi}%
  \BibitemOpen
  \bibfield  {author} {\bibinfo {author} {\bibfnamefont {Y.-T.}\ \bibnamefont
  {Tu}}, \bibinfo {author} {\bibfnamefont {Y.-C.}\ \bibnamefont {Tzeng}},\ and\
  \bibinfo {author} {\bibfnamefont {P.-Y.}\ \bibnamefont {Chang}},\ }\bibfield
  {title} {\bibinfo {title} {R{\'e}nyi entropies and negative central charges
  in non-hermitian quantum systems},\ }\href
  {https://doi.org/10.21468/SciPostPhys.12.6.194} {\bibfield  {journal}
  {\bibinfo  {journal} {SciPost Physics}\ }\textbf {\bibinfo {volume} {12}},\
  \bibinfo {pages} {194} (\bibinfo {year} {2022})}\BibitemShut {NoStop}%
\bibitem [{\citenamefont {Agarwal}\ \emph {et~al.}(2023)\citenamefont
  {Agarwal}, \citenamefont {Konar}, \citenamefont {Lakkaraju},\ and\
  \citenamefont {De}}]{agarwal2023recognizing}%
  \BibitemOpen
  \bibfield  {author} {\bibinfo {author} {\bibfnamefont {K.~D.}\ \bibnamefont
  {Agarwal}}, \bibinfo {author} {\bibfnamefont {T.~K.}\ \bibnamefont {Konar}},
  \bibinfo {author} {\bibfnamefont {L.~G.~C.}\ \bibnamefont {Lakkaraju}},\ and\
  \bibinfo {author} {\bibfnamefont {A.~S.}\ \bibnamefont {De}},\ }\bibfield
  {title} {\bibinfo {title} {Recognizing critical lines via entanglement in
  non-hermitian systems},\ }\href {https://arxiv.org/abs/2305.08374} {\bibfield
   {journal} {\bibinfo  {journal} {arXiv:2305.08374}\ } (\bibinfo {year}
  {2023})}\BibitemShut {NoStop}%
\bibitem [{\citenamefont {Xi}\ \emph {et~al.}(2021)\citenamefont {Xi},
  \citenamefont {Zhang}, \citenamefont {Gu},\ and\ \citenamefont
  {Chen}}]{xi2021classification}%
  \BibitemOpen
  \bibfield  {author} {\bibinfo {author} {\bibfnamefont {W.}~\bibnamefont
  {Xi}}, \bibinfo {author} {\bibfnamefont {Z.-H.}\ \bibnamefont {Zhang}},
  \bibinfo {author} {\bibfnamefont {Z.-C.}\ \bibnamefont {Gu}},\ and\ \bibinfo
  {author} {\bibfnamefont {W.-Q.}\ \bibnamefont {Chen}},\ }\bibfield  {title}
  {\bibinfo {title} {Classification of topological phases in one dimensional
  interacting non-hermitian systems and emergent unitarity},\ }\href
  {https://doi.org/https://doi.org/10.1016/j.scib.2021.04.027} {\bibfield
  {journal} {\bibinfo  {journal} {Science Bulletin}\ }\textbf {\bibinfo
  {volume} {66}},\ \bibinfo {pages} {1731} (\bibinfo {year}
  {2021})}\BibitemShut {NoStop}%
\bibitem [{\citenamefont {Lee}(2016)}]{lee2016anomalous}%
  \BibitemOpen
  \bibfield  {author} {\bibinfo {author} {\bibfnamefont {T.~E.}\ \bibnamefont
  {Lee}},\ }\bibfield  {title} {\bibinfo {title} {Anomalous edge state in a
  non-hermitian lattice},\ }\href
  {https://doi.org/https://doi.org/10.1103/PhysRevLett.116.133903} {\bibfield
  {journal} {\bibinfo  {journal} {Physical Review Letters}\ }\textbf {\bibinfo
  {volume} {116}},\ \bibinfo {pages} {133903} (\bibinfo {year}
  {2016})}\BibitemShut {NoStop}%
\bibitem [{\citenamefont {Yao}\ and\ \citenamefont {Wang}(2018)}]{yao2018edge}%
  \BibitemOpen
  \bibfield  {author} {\bibinfo {author} {\bibfnamefont {S.}~\bibnamefont
  {Yao}}\ and\ \bibinfo {author} {\bibfnamefont {Z.}~\bibnamefont {Wang}},\
  }\bibfield  {title} {\bibinfo {title} {Edge states and topological invariants
  of non-hermitian systems},\ }\href
  {https://doi.org/https://doi.org/10.1103/PhysRevLett.121.086803} {\bibfield
  {journal} {\bibinfo  {journal} {Physical Review Letters}\ }\textbf {\bibinfo
  {volume} {121}},\ \bibinfo {pages} {086803} (\bibinfo {year}
  {2018})}\BibitemShut {NoStop}%
\bibitem [{\citenamefont {Kunst}\ \emph {et~al.}(2018)\citenamefont {Kunst},
  \citenamefont {Edvardsson}, \citenamefont {Budich},\ and\ \citenamefont
  {Bergholtz}}]{kunst2018biorthogonal}%
  \BibitemOpen
  \bibfield  {author} {\bibinfo {author} {\bibfnamefont {F.~K.}\ \bibnamefont
  {Kunst}}, \bibinfo {author} {\bibfnamefont {E.}~\bibnamefont {Edvardsson}},
  \bibinfo {author} {\bibfnamefont {J.~C.}\ \bibnamefont {Budich}},\ and\
  \bibinfo {author} {\bibfnamefont {E.~J.}\ \bibnamefont {Bergholtz}},\
  }\bibfield  {title} {\bibinfo {title} {Biorthogonal bulk-boundary
  correspondence in non-hermitian systems},\ }\href
  {https://doi.org/https://doi.org/10.1103/PhysRevLett.121.026808} {\bibfield
  {journal} {\bibinfo  {journal} {Physical Review Letters}\ }\textbf {\bibinfo
  {volume} {121}},\ \bibinfo {pages} {026808} (\bibinfo {year}
  {2018})}\BibitemShut {NoStop}%
\bibitem [{\citenamefont {Xiong}(2018)}]{xiong2018does}%
  \BibitemOpen
  \bibfield  {author} {\bibinfo {author} {\bibfnamefont {Y.}~\bibnamefont
  {Xiong}},\ }\bibfield  {title} {\bibinfo {title} {Why does bulk boundary
  correspondence fail in some non-hermitian topological models},\ }\href
  {https://doi.org/https://doi.org/10.1088/2399-6528/aab64a} {\bibfield
  {journal} {\bibinfo  {journal} {Journal of Physics Communications}\ }\textbf
  {\bibinfo {volume} {2}},\ \bibinfo {pages} {035043} (\bibinfo {year}
  {2018})}\BibitemShut {NoStop}%
\bibitem [{\citenamefont {Gong}\ \emph {et~al.}(2018)\citenamefont {Gong},
  \citenamefont {Ashida}, \citenamefont {Kawabata}, \citenamefont {Takasan},
  \citenamefont {Higashikawa},\ and\ \citenamefont
  {Ueda}}]{gong2018topological}%
  \BibitemOpen
  \bibfield  {author} {\bibinfo {author} {\bibfnamefont {Z.}~\bibnamefont
  {Gong}}, \bibinfo {author} {\bibfnamefont {Y.}~\bibnamefont {Ashida}},
  \bibinfo {author} {\bibfnamefont {K.}~\bibnamefont {Kawabata}}, \bibinfo
  {author} {\bibfnamefont {K.}~\bibnamefont {Takasan}}, \bibinfo {author}
  {\bibfnamefont {S.}~\bibnamefont {Higashikawa}},\ and\ \bibinfo {author}
  {\bibfnamefont {M.}~\bibnamefont {Ueda}},\ }\bibfield  {title} {\bibinfo
  {title} {Topological phases of non-hermitian systems},\ }\href
  {https://doi.org/https://doi.org/10.1103/PhysRevX.8.031079} {\bibfield
  {journal} {\bibinfo  {journal} {Physical Review X}\ }\textbf {\bibinfo
  {volume} {8}},\ \bibinfo {pages} {031079} (\bibinfo {year}
  {2018})}\BibitemShut {NoStop}%
\bibitem [{\citenamefont {Martinez~Alvarez}\ \emph {et~al.}(2018)\citenamefont
  {Martinez~Alvarez}, \citenamefont {Barrios~Vargas},\ and\ \citenamefont
  {Foa~Torres}}]{alvarez2018non}%
  \BibitemOpen
  \bibfield  {author} {\bibinfo {author} {\bibfnamefont {V.~M.}\ \bibnamefont
  {Martinez~Alvarez}}, \bibinfo {author} {\bibfnamefont {J.~E.}\ \bibnamefont
  {Barrios~Vargas}},\ and\ \bibinfo {author} {\bibfnamefont {L.~E.~F.}\
  \bibnamefont {Foa~Torres}},\ }\bibfield  {title} {\bibinfo {title}
  {Non-hermitian robust edge states in one dimension: Anomalous localization
  and eigenspace condensation at exceptional points},\ }\href
  {https://doi.org/https://doi.org/10.1103/PhysRevB.97.121401} {\bibfield
  {journal} {\bibinfo  {journal} {Physical Review B}\ }\textbf {\bibinfo
  {volume} {97}},\ \bibinfo {pages} {121401(R)} (\bibinfo {year}
  {2018})}\BibitemShut {NoStop}%
\bibitem [{\citenamefont {Yokomizo}\ and\ \citenamefont
  {Murakami}(2019)}]{yokomizo2019non}%
  \BibitemOpen
  \bibfield  {author} {\bibinfo {author} {\bibfnamefont {K.}~\bibnamefont
  {Yokomizo}}\ and\ \bibinfo {author} {\bibfnamefont {S.}~\bibnamefont
  {Murakami}},\ }\bibfield  {title} {\bibinfo {title} {Non-bloch band theory of
  non-hermitian systems},\ }\href
  {https://doi.org/https://doi.org/10.1103/PhysRevLett.123.066404} {\bibfield
  {journal} {\bibinfo  {journal} {Physical Review Letters}\ }\textbf {\bibinfo
  {volume} {123}},\ \bibinfo {pages} {066404} (\bibinfo {year}
  {2019})}\BibitemShut {NoStop}%
\bibitem [{\citenamefont {Okuma}\ \emph {et~al.}(2020)\citenamefont {Okuma},
  \citenamefont {Kawabata}, \citenamefont {Shiozaki},\ and\ \citenamefont
  {Sato}}]{okuma2020topological}%
  \BibitemOpen
  \bibfield  {author} {\bibinfo {author} {\bibfnamefont {N.}~\bibnamefont
  {Okuma}}, \bibinfo {author} {\bibfnamefont {K.}~\bibnamefont {Kawabata}},
  \bibinfo {author} {\bibfnamefont {K.}~\bibnamefont {Shiozaki}},\ and\
  \bibinfo {author} {\bibfnamefont {M.}~\bibnamefont {Sato}},\ }\bibfield
  {title} {\bibinfo {title} {Topological origin of non-hermitian skin
  effects},\ }\href
  {https://doi.org/https://doi.org/10.1103/PhysRevLett.124.086801} {\bibfield
  {journal} {\bibinfo  {journal} {Physical Review Letters}\ }\textbf {\bibinfo
  {volume} {124}},\ \bibinfo {pages} {086801} (\bibinfo {year}
  {2020})}\BibitemShut {NoStop}%
\bibitem [{\citenamefont {Zhang}\ \emph {et~al.}(2020)\citenamefont {Zhang},
  \citenamefont {Yang},\ and\ \citenamefont {Fang}}]{zhang2020correspondence}%
  \BibitemOpen
  \bibfield  {author} {\bibinfo {author} {\bibfnamefont {K.}~\bibnamefont
  {Zhang}}, \bibinfo {author} {\bibfnamefont {Z.}~\bibnamefont {Yang}},\ and\
  \bibinfo {author} {\bibfnamefont {C.}~\bibnamefont {Fang}},\ }\bibfield
  {title} {\bibinfo {title} {Correspondence between winding numbers and skin
  modes in non-hermitian systems},\ }\href
  {https://doi.org/https://doi.org/10.1103/PhysRevLett.125.126402} {\bibfield
  {journal} {\bibinfo  {journal} {Physical Review Letters}\ }\textbf {\bibinfo
  {volume} {125}},\ \bibinfo {pages} {126402} (\bibinfo {year}
  {2020})}\BibitemShut {NoStop}%
\bibitem [{\citenamefont {Yang}\ \emph {et~al.}(2020)\citenamefont {Yang},
  \citenamefont {Zhang}, \citenamefont {Fang},\ and\ \citenamefont
  {Hu}}]{yang2020non}%
  \BibitemOpen
  \bibfield  {author} {\bibinfo {author} {\bibfnamefont {Z.}~\bibnamefont
  {Yang}}, \bibinfo {author} {\bibfnamefont {K.}~\bibnamefont {Zhang}},
  \bibinfo {author} {\bibfnamefont {C.}~\bibnamefont {Fang}},\ and\ \bibinfo
  {author} {\bibfnamefont {J.}~\bibnamefont {Hu}},\ }\bibfield  {title}
  {\bibinfo {title} {Non-hermitian bulk-boundary correspondence and auxiliary
  generalized brillouin zone theory},\ }\href
  {https://doi.org/https://doi.org/10.1103/PhysRevLett.125.226402} {\bibfield
  {journal} {\bibinfo  {journal} {Physical Review Letters}\ }\textbf {\bibinfo
  {volume} {125}},\ \bibinfo {pages} {226402} (\bibinfo {year}
  {2020})}\BibitemShut {NoStop}%
\bibitem [{\citenamefont {Wang}\ \emph {et~al.}(2020)\citenamefont {Wang},
  \citenamefont {Guo},\ and\ \citenamefont {Kou}}]{wang2020defective}%
  \BibitemOpen
  \bibfield  {author} {\bibinfo {author} {\bibfnamefont {X.-R.}\ \bibnamefont
  {Wang}}, \bibinfo {author} {\bibfnamefont {C.-X.}\ \bibnamefont {Guo}},\ and\
  \bibinfo {author} {\bibfnamefont {S.-P.}\ \bibnamefont {Kou}},\ }\bibfield
  {title} {\bibinfo {title} {Defective edge states and number-anomalous
  bulk-boundary correspondence in non-hermitian topological systems},\ }\href
  {https://doi.org/https://doi.org/10.1103/PhysRevB.101.121116} {\bibfield
  {journal} {\bibinfo  {journal} {Physical Review B}\ }\textbf {\bibinfo
  {volume} {101}},\ \bibinfo {pages} {121116(R)} (\bibinfo {year}
  {2020})}\BibitemShut {NoStop}%
\bibitem [{\citenamefont {Jiang}\ \emph {et~al.}(2020)\citenamefont {Jiang},
  \citenamefont {L{\"u}},\ and\ \citenamefont {Chen}}]{jiang2020topological}%
  \BibitemOpen
  \bibfield  {author} {\bibinfo {author} {\bibfnamefont {H.}~\bibnamefont
  {Jiang}}, \bibinfo {author} {\bibfnamefont {R.}~\bibnamefont {L{\"u}}},\ and\
  \bibinfo {author} {\bibfnamefont {S.}~\bibnamefont {Chen}},\ }\bibfield
  {title} {\bibinfo {title} {Topological invariants, zero mode edge states and
  finite size effect for a generalized non-reciprocal su-schrieffer-heeger
  model},\ }\href {https://doi.org/https://doi.org/10.1140/epjb/e2020-10036-3}
  {\bibfield  {journal} {\bibinfo  {journal} {The European Physical Journal B}\
  }\textbf {\bibinfo {volume} {93}},\ \bibinfo {pages} {1} (\bibinfo {year}
  {2020})}\BibitemShut {NoStop}%
\bibitem [{\citenamefont {Weidemann}\ \emph {et~al.}(2020)\citenamefont
  {Weidemann}, \citenamefont {Kremer}, \citenamefont {Helbig}, \citenamefont
  {Hofmann}, \citenamefont {Stegmaier}, \citenamefont {Greiter}, \citenamefont
  {Thomale},\ and\ \citenamefont {Szameit}}]{weidemann2020topological}%
  \BibitemOpen
  \bibfield  {author} {\bibinfo {author} {\bibfnamefont {S.}~\bibnamefont
  {Weidemann}}, \bibinfo {author} {\bibfnamefont {M.}~\bibnamefont {Kremer}},
  \bibinfo {author} {\bibfnamefont {T.}~\bibnamefont {Helbig}}, \bibinfo
  {author} {\bibfnamefont {T.}~\bibnamefont {Hofmann}}, \bibinfo {author}
  {\bibfnamefont {A.}~\bibnamefont {Stegmaier}}, \bibinfo {author}
  {\bibfnamefont {M.}~\bibnamefont {Greiter}}, \bibinfo {author} {\bibfnamefont
  {R.}~\bibnamefont {Thomale}},\ and\ \bibinfo {author} {\bibfnamefont
  {A.}~\bibnamefont {Szameit}},\ }\bibfield  {title} {\bibinfo {title}
  {Topological funneling of light},\ }\href
  {https://doi.org/10.1126/science.aaz8727} {\bibfield  {journal} {\bibinfo
  {journal} {Science}\ }\textbf {\bibinfo {volume} {368}},\ \bibinfo {pages}
  {311} (\bibinfo {year} {2020})}\BibitemShut {NoStop}%
\bibitem [{\citenamefont {Xiao}\ \emph {et~al.}(2020)\citenamefont {Xiao},
  \citenamefont {Deng}, \citenamefont {Wang}, \citenamefont {Zhu},
  \citenamefont {Wang}, \citenamefont {Yi},\ and\ \citenamefont
  {Xue}}]{xiao2020non}%
  \BibitemOpen
  \bibfield  {author} {\bibinfo {author} {\bibfnamefont {L.}~\bibnamefont
  {Xiao}}, \bibinfo {author} {\bibfnamefont {T.}~\bibnamefont {Deng}}, \bibinfo
  {author} {\bibfnamefont {K.}~\bibnamefont {Wang}}, \bibinfo {author}
  {\bibfnamefont {G.}~\bibnamefont {Zhu}}, \bibinfo {author} {\bibfnamefont
  {Z.}~\bibnamefont {Wang}}, \bibinfo {author} {\bibfnamefont {W.}~\bibnamefont
  {Yi}},\ and\ \bibinfo {author} {\bibfnamefont {P.}~\bibnamefont {Xue}},\
  }\bibfield  {title} {\bibinfo {title} {Non-hermitian bulk--boundary
  correspondence in quantum dynamics},\ }\href
  {https://doi.org/https://doi.org/10.1038/s41567-020-0836-6} {\bibfield
  {journal} {\bibinfo  {journal} {Nature Physics}\ }\textbf {\bibinfo {volume}
  {16}},\ \bibinfo {pages} {761} (\bibinfo {year} {2020})}\BibitemShut
  {NoStop}%
\bibitem [{\citenamefont {Borgnia}\ \emph {et~al.}(2020)\citenamefont
  {Borgnia}, \citenamefont {Kruchkov},\ and\ \citenamefont
  {Slager}}]{borgnia2020non}%
  \BibitemOpen
  \bibfield  {author} {\bibinfo {author} {\bibfnamefont {D.~S.}\ \bibnamefont
  {Borgnia}}, \bibinfo {author} {\bibfnamefont {A.~J.}\ \bibnamefont
  {Kruchkov}},\ and\ \bibinfo {author} {\bibfnamefont {R.-J.}\ \bibnamefont
  {Slager}},\ }\bibfield  {title} {\bibinfo {title} {Non-hermitian boundary
  modes and topology},\ }\href
  {https://doi.org/https://doi.org/10.1103/PhysRevLett.124.056802} {\bibfield
  {journal} {\bibinfo  {journal} {Physical Review Letters}\ }\textbf {\bibinfo
  {volume} {124}},\ \bibinfo {pages} {056802} (\bibinfo {year}
  {2020})}\BibitemShut {NoStop}%
\bibitem [{\citenamefont {Heiss}(2012)}]{heiss2012physics}%
  \BibitemOpen
  \bibfield  {author} {\bibinfo {author} {\bibfnamefont {W.}~\bibnamefont
  {Heiss}},\ }\bibfield  {title} {\bibinfo {title} {The physics of exceptional
  points},\ }\href
  {https://doi.org/https://doi.org/10.1088/1751-8113/45/44/444016} {\bibfield
  {journal} {\bibinfo  {journal} {Journal of Physics A: Mathematical and
  Theoretical}\ }\textbf {\bibinfo {volume} {45}},\ \bibinfo {pages} {444016}
  (\bibinfo {year} {2012})}\BibitemShut {NoStop}%
\bibitem [{\citenamefont {Kozii}\ and\ \citenamefont
  {Fu}(2024)}]{kozii2024non}%
  \BibitemOpen
  \bibfield  {author} {\bibinfo {author} {\bibfnamefont {V.}~\bibnamefont
  {Kozii}}\ and\ \bibinfo {author} {\bibfnamefont {L.}~\bibnamefont {Fu}},\
  }\bibfield  {title} {\bibinfo {title} {Non-hermitian topological theory of
  finite-lifetime quasiparticles: Prediction of bulk fermi arc due to
  exceptional point},\ }\href
  {https://doi.org/https://doi.org/10.1103/PhysRevB.109.235139} {\bibfield
  {journal} {\bibinfo  {journal} {Physical Review B}\ }\textbf {\bibinfo
  {volume} {109}},\ \bibinfo {pages} {235139} (\bibinfo {year}
  {2024})}\BibitemShut {NoStop}%
\bibitem [{\citenamefont {Hodaei}\ \emph {et~al.}(2017)\citenamefont {Hodaei},
  \citenamefont {Hassan}, \citenamefont {Wittek}, \citenamefont
  {Garcia-Gracia}, \citenamefont {El-Ganainy}, \citenamefont
  {Christodoulides},\ and\ \citenamefont {Khajavikhan}}]{hodaei2017enhanced}%
  \BibitemOpen
  \bibfield  {author} {\bibinfo {author} {\bibfnamefont {H.}~\bibnamefont
  {Hodaei}}, \bibinfo {author} {\bibfnamefont {A.~U.}\ \bibnamefont {Hassan}},
  \bibinfo {author} {\bibfnamefont {S.}~\bibnamefont {Wittek}}, \bibinfo
  {author} {\bibfnamefont {H.}~\bibnamefont {Garcia-Gracia}}, \bibinfo {author}
  {\bibfnamefont {R.}~\bibnamefont {El-Ganainy}}, \bibinfo {author}
  {\bibfnamefont {D.~N.}\ \bibnamefont {Christodoulides}},\ and\ \bibinfo
  {author} {\bibfnamefont {M.}~\bibnamefont {Khajavikhan}},\ }\bibfield
  {title} {\bibinfo {title} {Enhanced sensitivity at higher-order exceptional
  points},\ }\href {https://doi.org/https://doi.org/10.1038/nature23280}
  {\bibfield  {journal} {\bibinfo  {journal} {Nature}\ }\textbf {\bibinfo
  {volume} {548}},\ \bibinfo {pages} {187} (\bibinfo {year}
  {2017})}\BibitemShut {NoStop}%
\bibitem [{\citenamefont {Zhou}\ \emph {et~al.}(2018)\citenamefont {Zhou},
  \citenamefont {Peng}, \citenamefont {Yoon}, \citenamefont {Hsu},
  \citenamefont {Nelson}, \citenamefont {Fu}, \citenamefont {Joannopoulos},
  \citenamefont {Solja{\v{c}}i{\'c}},\ and\ \citenamefont
  {Zhen}}]{zhou2018observation}%
  \BibitemOpen
  \bibfield  {author} {\bibinfo {author} {\bibfnamefont {H.}~\bibnamefont
  {Zhou}}, \bibinfo {author} {\bibfnamefont {C.}~\bibnamefont {Peng}}, \bibinfo
  {author} {\bibfnamefont {Y.}~\bibnamefont {Yoon}}, \bibinfo {author}
  {\bibfnamefont {C.~W.}\ \bibnamefont {Hsu}}, \bibinfo {author} {\bibfnamefont
  {K.~A.}\ \bibnamefont {Nelson}}, \bibinfo {author} {\bibfnamefont
  {L.}~\bibnamefont {Fu}}, \bibinfo {author} {\bibfnamefont {J.~D.}\
  \bibnamefont {Joannopoulos}}, \bibinfo {author} {\bibfnamefont
  {M.}~\bibnamefont {Solja{\v{c}}i{\'c}}},\ and\ \bibinfo {author}
  {\bibfnamefont {B.}~\bibnamefont {Zhen}},\ }\bibfield  {title} {\bibinfo
  {title} {Observation of bulk fermi arc and polarization half charge from
  paired exceptional points},\ }\href {https://doi.org/10.1126/science.aap9859}
  {\bibfield  {journal} {\bibinfo  {journal} {Science}\ }\textbf {\bibinfo
  {volume} {359}},\ \bibinfo {pages} {1009} (\bibinfo {year}
  {2018})}\BibitemShut {NoStop}%
\bibitem [{\citenamefont {Miri}\ and\ \citenamefont
  {Alu}(2019)}]{miri2019exceptional}%
  \BibitemOpen
  \bibfield  {author} {\bibinfo {author} {\bibfnamefont {M.-A.}\ \bibnamefont
  {Miri}}\ and\ \bibinfo {author} {\bibfnamefont {A.}~\bibnamefont {Alu}},\
  }\bibfield  {title} {\bibinfo {title} {Exceptional points in optics and
  photonics},\ }\href {https://doi.org/10.1126/science.aar7709} {\bibfield
  {journal} {\bibinfo  {journal} {Science}\ }\textbf {\bibinfo {volume}
  {363}},\ \bibinfo {pages} {42} (\bibinfo {year} {2019})}\BibitemShut
  {NoStop}%
\bibitem [{\citenamefont {Park}\ \emph {et~al.}(2019)\citenamefont {Park},
  \citenamefont {Ndao}, \citenamefont {Cai}, \citenamefont {Hsu}, \citenamefont
  {Kodigala}, \citenamefont {Lepetit}, \citenamefont {Lo},\ and\ \citenamefont
  {Kant{\'e}}}]{park2019observation}%
  \BibitemOpen
  \bibfield  {author} {\bibinfo {author} {\bibfnamefont {J.-H.}\ \bibnamefont
  {Park}}, \bibinfo {author} {\bibfnamefont {A.}~\bibnamefont {Ndao}}, \bibinfo
  {author} {\bibfnamefont {W.}~\bibnamefont {Cai}}, \bibinfo {author}
  {\bibfnamefont {L.-Y.}\ \bibnamefont {Hsu}}, \bibinfo {author} {\bibfnamefont
  {A.}~\bibnamefont {Kodigala}}, \bibinfo {author} {\bibfnamefont
  {T.}~\bibnamefont {Lepetit}}, \bibinfo {author} {\bibfnamefont {Y.-H.}\
  \bibnamefont {Lo}},\ and\ \bibinfo {author} {\bibfnamefont {B.}~\bibnamefont
  {Kant{\'e}}},\ }\bibfield  {title} {\bibinfo {title} {Observation of
  plasmonic exceptional points},\ }\href {https://arxiv.org/abs/1904.01073}
  {\bibfield  {journal} {\bibinfo  {journal} {arXiv preprint arXiv:1904.01073}\
  } (\bibinfo {year} {2019})}\BibitemShut {NoStop}%
\bibitem [{\citenamefont {Yang}\ and\ \citenamefont {Hu}(2019)}]{yang2019non}%
  \BibitemOpen
  \bibfield  {author} {\bibinfo {author} {\bibfnamefont {Z.}~\bibnamefont
  {Yang}}\ and\ \bibinfo {author} {\bibfnamefont {J.}~\bibnamefont {Hu}},\
  }\bibfield  {title} {\bibinfo {title} {Non-hermitian hopf-link exceptional
  line semimetals},\ }\href
  {https://doi.org/https://doi.org/10.1103/PhysRevB.99.081102} {\bibfield
  {journal} {\bibinfo  {journal} {Physical Review B}\ }\textbf {\bibinfo
  {volume} {99}},\ \bibinfo {pages} {081102(R)} (\bibinfo {year}
  {2019})}\BibitemShut {NoStop}%
\bibitem [{\citenamefont {{\"O}zdemir}\ \emph {et~al.}(2019)\citenamefont
  {{\"O}zdemir}, \citenamefont {Rotter}, \citenamefont {Nori},\ and\
  \citenamefont {Yang}}]{ozdemir2019parity}%
  \BibitemOpen
  \bibfield  {author} {\bibinfo {author} {\bibfnamefont {{\c{S}}.}~\bibnamefont
  {{\"O}zdemir}}, \bibinfo {author} {\bibfnamefont {S.}~\bibnamefont {Rotter}},
  \bibinfo {author} {\bibfnamefont {F.}~\bibnamefont {Nori}},\ and\ \bibinfo
  {author} {\bibfnamefont {L.}~\bibnamefont {Yang}},\ }\bibfield  {title}
  {\bibinfo {title} {Parity-time symmetry and exceptional points in
  photonics},\ }\href
  {https://doi.org/https://doi.org/10.1038/s41563-019-0304-9} {\bibfield
  {journal} {\bibinfo  {journal} {Nature Materials}\ }\textbf {\bibinfo
  {volume} {18}},\ \bibinfo {pages} {783} (\bibinfo {year} {2019})}\BibitemShut
  {NoStop}%
\bibitem [{\citenamefont {D{\'o}ra}\ \emph {et~al.}(2019)\citenamefont
  {D{\'o}ra}, \citenamefont {Heyl},\ and\ \citenamefont
  {Moessner}}]{dora2019kibble}%
  \BibitemOpen
  \bibfield  {author} {\bibinfo {author} {\bibfnamefont {B.}~\bibnamefont
  {D{\'o}ra}}, \bibinfo {author} {\bibfnamefont {M.}~\bibnamefont {Heyl}},\
  and\ \bibinfo {author} {\bibfnamefont {R.}~\bibnamefont {Moessner}},\
  }\bibfield  {title} {\bibinfo {title} {The kibble-zurek mechanism at
  exceptional points},\ }\href
  {https://doi.org/https://doi.org/10.1038/s41467-019-10048-9} {\bibfield
  {journal} {\bibinfo  {journal} {Nature Communications}\ }\textbf {\bibinfo
  {volume} {10}},\ \bibinfo {pages} {1} (\bibinfo {year} {2019})}\BibitemShut
  {NoStop}%
\bibitem [{\citenamefont {Jin}\ \emph {et~al.}(2020)\citenamefont {Jin},
  \citenamefont {Wu}, \citenamefont {Wei},\ and\ \citenamefont
  {Song}}]{jin2020hybrid}%
  \BibitemOpen
  \bibfield  {author} {\bibinfo {author} {\bibfnamefont {L.}~\bibnamefont
  {Jin}}, \bibinfo {author} {\bibfnamefont {H.~C.}\ \bibnamefont {Wu}},
  \bibinfo {author} {\bibfnamefont {B.-B.}\ \bibnamefont {Wei}},\ and\ \bibinfo
  {author} {\bibfnamefont {Z.}~\bibnamefont {Song}},\ }\bibfield  {title}
  {\bibinfo {title} {Hybrid exceptional point created from type-iii dirac
  point},\ }\href {https://doi.org/https://doi.org/10.1103/PhysRevB.101.045130}
  {\bibfield  {journal} {\bibinfo  {journal} {Physical Review B}\ }\textbf
  {\bibinfo {volume} {101}},\ \bibinfo {pages} {045130} (\bibinfo {year}
  {2020})}\BibitemShut {NoStop}%
\bibitem [{\citenamefont {Xiao}\ \emph {et~al.}(2021)\citenamefont {Xiao},
  \citenamefont {Deng}, \citenamefont {Wang}, \citenamefont {Wang},
  \citenamefont {Yi},\ and\ \citenamefont {Xue}}]{xiao2021observation}%
  \BibitemOpen
  \bibfield  {author} {\bibinfo {author} {\bibfnamefont {L.}~\bibnamefont
  {Xiao}}, \bibinfo {author} {\bibfnamefont {T.}~\bibnamefont {Deng}}, \bibinfo
  {author} {\bibfnamefont {K.}~\bibnamefont {Wang}}, \bibinfo {author}
  {\bibfnamefont {Z.}~\bibnamefont {Wang}}, \bibinfo {author} {\bibfnamefont
  {W.}~\bibnamefont {Yi}},\ and\ \bibinfo {author} {\bibfnamefont
  {P.}~\bibnamefont {Xue}},\ }\bibfield  {title} {\bibinfo {title} {Observation
  of non-bloch parity-time symmetry and exceptional points},\ }\href
  {https://doi.org/https://doi.org/10.1103/PhysRevLett.126.230402} {\bibfield
  {journal} {\bibinfo  {journal} {Physical Review Letters}\ }\textbf {\bibinfo
  {volume} {126}},\ \bibinfo {pages} {230402} (\bibinfo {year}
  {2021})}\BibitemShut {NoStop}%
\bibitem [{\citenamefont {Chen}\ \emph
  {et~al.}(2022{\natexlab{b}})\citenamefont {Chen}, \citenamefont {Liu},
  \citenamefont {Zhao}, \citenamefont {Hu},\ and\ \citenamefont
  {Fu}}]{chen2022asymmetric}%
  \BibitemOpen
  \bibfield  {author} {\bibinfo {author} {\bibfnamefont {C.}~\bibnamefont
  {Chen}}, \bibinfo {author} {\bibfnamefont {Y.}~\bibnamefont {Liu}}, \bibinfo
  {author} {\bibfnamefont {L.}~\bibnamefont {Zhao}}, \bibinfo {author}
  {\bibfnamefont {X.}~\bibnamefont {Hu}},\ and\ \bibinfo {author}
  {\bibfnamefont {Y.}~\bibnamefont {Fu}},\ }\bibfield  {title} {\bibinfo
  {title} {Asymmetric nonlinear-mode-conversion in an optical waveguide with
  $\mathcal{PT}$ symmetry},\ }\href
  {https://doi.org/https://doi.org/10.1007/s11467-022-1177-y} {\bibfield
  {journal} {\bibinfo  {journal} {Frontiers of Physics}\ }\textbf {\bibinfo
  {volume} {17}},\ \bibinfo {pages} {52504} (\bibinfo {year}
  {2022}{\natexlab{b}})}\BibitemShut {NoStop}%
\bibitem [{\citenamefont {Li}\ \emph {et~al.}(2023)\citenamefont {Li},
  \citenamefont {Wang}, \citenamefont {Song},\ and\ \citenamefont
  {Wang}}]{li2023non}%
  \BibitemOpen
  \bibfield  {author} {\bibinfo {author} {\bibfnamefont {B.}~\bibnamefont
  {Li}}, \bibinfo {author} {\bibfnamefont {H.-R.}\ \bibnamefont {Wang}},
  \bibinfo {author} {\bibfnamefont {F.}~\bibnamefont {Song}},\ and\ \bibinfo
  {author} {\bibfnamefont {Z.}~\bibnamefont {Wang}},\ }\bibfield  {title}
  {\bibinfo {title} {Non-bloch dynamics and topology in a classical
  non-equilibrium process},\ }\href {https://arxiv.org/abs/2306.11105}
  {\bibfield  {journal} {\bibinfo  {journal} {arXiv preprint arXiv:2306.11105}\
  } (\bibinfo {year} {2023})}\BibitemShut {NoStop}%
\bibitem [{\citenamefont {Lee}\ and\ \citenamefont
  {Chan}(2014)}]{lee2014heralded}%
  \BibitemOpen
  \bibfield  {author} {\bibinfo {author} {\bibfnamefont {T.~E.}\ \bibnamefont
  {Lee}}\ and\ \bibinfo {author} {\bibfnamefont {C.-K.}\ \bibnamefont {Chan}},\
  }\bibfield  {title} {\bibinfo {title} {Heralded magnetism in non-hermitian
  atomic systems},\ }\href
  {https://doi.org/https://doi.org/10.1103/PhysRevX.4.041001} {\bibfield
  {journal} {\bibinfo  {journal} {Physical Review X}\ }\textbf {\bibinfo
  {volume} {4}},\ \bibinfo {pages} {041001} (\bibinfo {year}
  {2014})}\BibitemShut {NoStop}%
\bibitem [{\citenamefont {Zeng}\ \emph {et~al.}(2016)\citenamefont {Zeng},
  \citenamefont {Zhu}, \citenamefont {Chen}, \citenamefont {You},\ and\
  \citenamefont {L{\"u}}}]{zeng2016non}%
  \BibitemOpen
  \bibfield  {author} {\bibinfo {author} {\bibfnamefont {Q.-B.}\ \bibnamefont
  {Zeng}}, \bibinfo {author} {\bibfnamefont {B.}~\bibnamefont {Zhu}}, \bibinfo
  {author} {\bibfnamefont {S.}~\bibnamefont {Chen}}, \bibinfo {author}
  {\bibfnamefont {L.}~\bibnamefont {You}},\ and\ \bibinfo {author}
  {\bibfnamefont {R.}~\bibnamefont {L{\"u}}},\ }\bibfield  {title} {\bibinfo
  {title} {Non-hermitian kitaev chain with complex on-site potentials},\ }\href
  {https://doi.org/https://doi.org/10.1103/PhysRevA.94.022119} {\bibfield
  {journal} {\bibinfo  {journal} {Physical Review A}\ }\textbf {\bibinfo
  {volume} {94}},\ \bibinfo {pages} {022119} (\bibinfo {year}
  {2016})}\BibitemShut {NoStop}%
\bibitem [{\citenamefont {Biella}\ and\ \citenamefont
  {Schir{\'o}}(2021)}]{biella2021many}%
  \BibitemOpen
  \bibfield  {author} {\bibinfo {author} {\bibfnamefont {A.}~\bibnamefont
  {Biella}}\ and\ \bibinfo {author} {\bibfnamefont {M.}~\bibnamefont
  {Schir{\'o}}},\ }\bibfield  {title} {\bibinfo {title} {Many-body quantum zeno
  effect and measurement-induced subradiance transition},\ }\href
  {https://doi.org/https://doi.org/10.22331/q-2021-08-19-528} {\bibfield
  {journal} {\bibinfo  {journal} {Quantum}\ }\textbf {\bibinfo {volume} {5}},\
  \bibinfo {pages} {528} (\bibinfo {year} {2021})}\BibitemShut {NoStop}%
\bibitem [{\citenamefont {Turkeshi}\ and\ \citenamefont
  {Schir{\'o}}(2023)}]{turkeshi2023entanglement}%
  \BibitemOpen
  \bibfield  {author} {\bibinfo {author} {\bibfnamefont {X.}~\bibnamefont
  {Turkeshi}}\ and\ \bibinfo {author} {\bibfnamefont {M.}~\bibnamefont
  {Schir{\'o}}},\ }\bibfield  {title} {\bibinfo {title} {Entanglement and
  correlation spreading in non-hermitian spin chains},\ }\href
  {https://doi.org/https://doi.org/10.1103/PhysRevB.107.L020403} {\bibfield
  {journal} {\bibinfo  {journal} {Physical Review B}\ }\textbf {\bibinfo
  {volume} {107}},\ \bibinfo {pages} {L020403} (\bibinfo {year}
  {2023})}\BibitemShut {NoStop}%
\bibitem [{\citenamefont {Turkeshi}\ \emph {et~al.}(2021)\citenamefont
  {Turkeshi}, \citenamefont {Biella}, \citenamefont {Fazio}, \citenamefont
  {Dalmonte},\ and\ \citenamefont {Schir{\'o}}}]{turkeshi2021measurement}%
  \BibitemOpen
  \bibfield  {author} {\bibinfo {author} {\bibfnamefont {X.}~\bibnamefont
  {Turkeshi}}, \bibinfo {author} {\bibfnamefont {A.}~\bibnamefont {Biella}},
  \bibinfo {author} {\bibfnamefont {R.}~\bibnamefont {Fazio}}, \bibinfo
  {author} {\bibfnamefont {M.}~\bibnamefont {Dalmonte}},\ and\ \bibinfo
  {author} {\bibfnamefont {M.}~\bibnamefont {Schir{\'o}}},\ }\bibfield  {title}
  {\bibinfo {title} {Measurement-induced entanglement transitions in the
  quantum ising chain: From infinite to zero clicks},\ }\href
  {https://doi.org/https://doi.org/10.1103/PhysRevB.103.224210} {\bibfield
  {journal} {\bibinfo  {journal} {Physical Review B}\ }\textbf {\bibinfo
  {volume} {103}},\ \bibinfo {pages} {224210} (\bibinfo {year}
  {2021})}\BibitemShut {NoStop}%
\bibitem [{\citenamefont {Turkeshi}\ \emph {et~al.}(2022)\citenamefont
  {Turkeshi}, \citenamefont {Dalmonte}, \citenamefont {Fazio},\ and\
  \citenamefont {Schir{\`o}}}]{turkeshi2022entanglement}%
  \BibitemOpen
  \bibfield  {author} {\bibinfo {author} {\bibfnamefont {X.}~\bibnamefont
  {Turkeshi}}, \bibinfo {author} {\bibfnamefont {M.}~\bibnamefont {Dalmonte}},
  \bibinfo {author} {\bibfnamefont {R.}~\bibnamefont {Fazio}},\ and\ \bibinfo
  {author} {\bibfnamefont {M.}~\bibnamefont {Schir{\`o}}},\ }\bibfield  {title}
  {\bibinfo {title} {Entanglement transitions from stochastic resetting of
  non-hermitian quasiparticles},\ }\href
  {https://doi.org/https://doi.org/10.1103/PhysRevB.105.L241114} {\bibfield
  {journal} {\bibinfo  {journal} {Physical Review B}\ }\textbf {\bibinfo
  {volume} {105}},\ \bibinfo {pages} {L241114} (\bibinfo {year}
  {2022})}\BibitemShut {NoStop}%
\bibitem [{\citenamefont {Sun}\ and\ \citenamefont
  {Kou}(2023)}]{sun2023aufbau}%
  \BibitemOpen
  \bibfield  {author} {\bibinfo {author} {\bibfnamefont {G.}~\bibnamefont
  {Sun}}\ and\ \bibinfo {author} {\bibfnamefont {S.-P.}\ \bibnamefont {Kou}},\
  }\bibfield  {title} {\bibinfo {title} {Aufbau principle for non-hermitian
  systems},\ }\href {https://arxiv.org/abs/2307.04696} {\bibfield  {journal}
  {\bibinfo  {journal} {arXiv:2307.04696}\ } (\bibinfo {year}
  {2023})}\BibitemShut {NoStop}%
\end{thebibliography}%

\end{document}